\documentclass[graybox]{svmult}
\pdfoutput=1

\usepackage{type1cm}        % activate if the above 3 fonts are
\usepackage{makeidx}         % allows index generation
\usepackage{graphicx}        % standard LaTeX graphics tool
\usepackage{multicol}        % used for the two-column index
\usepackage[bottom]{footmisc}% places footnotes at page bottom
\usepackage{cite}  % to group multiple citations

\usepackage{newtxtext}       % 
\usepackage[varvw]{newtxmath}       % selects Times Roman as basic font

\usepackage{hyperref}

\newcommand{\be}{\begin{equation}}
\newcommand{\ee}{\end{equation}} 

\newcommand{\f}{\frac}
\newcommand{\p}{\partial}
\newcommand{\la}{\langle}
\newcommand{\ra}{\rangle}

\newcommand{\extd}{\textrm{d}}
\DeclareMathOperator{\Tr}{Tr}

\let\a=\alpha \let\b=\beta    \let\d=\delta  
        \let\l=\lambda
               \let\om=\omega
\let\s=\sigma \let\t=\tau     
    
 \let\D=\Delta   \let\L=\Lambda \let\X=F
         
  \let\eps=\epsilon

\newcommand{\cB}{\mathcal{B}}

\newcommand{\cH}{\mathcal{H}}

\newcommand{\cL}{\mathcal{L}}

\newcommand{\cN}{\mathcal{N}}
\newcommand{\cO}{\mathcal{O}}
\newcommand{\cP}{\mathcal{P}}

\newcommand{\cT}{\mathcal{T}}

\makeindex             % used for the subject index
\begin{document}

\title*{Landau Theory of Causal Dynamical Triangulations}
\author{Dario Benedetti}
\institute{Dario Benedetti \at  CPHT, CNRS, \'{E}cole polytechnique, Institut Polytechnique de Paris, 91120 Palaiseau, France. \email{dario.benedetti@polytechnique.edu}}
%
% Use th1e package "url.sty" to avoid
\maketitle

\abstract{Understanding the continuum limit of a theory of discrete random geometries is a beautiful but difficult challenge. In this optic, we review here the insights that can be obtained for Causal Dynamical Triangulations (CDT) by employing the Landau approach to critical phenomena. In particular, concentrating on the cases of two and three dimensions, we will make the case that the configuration of the volume of spatial slices effectively plays the role of an order parameter, helping us to understand the phase structure of CDT. Moreover, consistency with numerical simulations of CDT provides hints that the effective field theory of the model lives in the space of theories invariant under foliation-preserving diffeomorphisms. Among such theories, Ho\v{r}ava-Lifshitz gravity has the special status of being a perturbatively renormalizable theory, while General Relativity sits in a subspace with enhanced symmetry. In order to reach either of them, one would likely need to fine tune some of the parameters in the CDT action, or additional ones from some generalization thereof. }

\keywords{Quantum Gravity, Random Geometry, Lattice Models, Phase Diagrams, Causal Dynamical Triangulations, Ho\v{r}ava-Lifshitz Gravity}

\

\noindent {\bf Note:} This is a contribution to the \emph{Handbook of Quantum Gravity} which is expected to be published in 2023. It will appear as a chapter in the section dedicated to \emph{Causal Dynamical Triangulations}.  %only for arXiv

\newpage

%%%%%%%%%%%%%%%%%%%%%%%%%
\section{Introduction}
\label{sec:intro}
%%%%%%%%%%%%%%%%%%%%%%%%%%

Lattice regularization is a common approach to taming in a nonperturbative way the ultraviolet divergences of a quantum field theory. For many quantum field theories it is relatively straightforward to formulate a corresponding lattice field theory, but showing that a nontrivial continuum limit can be obtained is generally very difficult.\footnote{For example, only recently is has been rigorously proven that the continumm limit of the $\phi^4$ theory in four dimensions is trivial (i.e.\ non-interacting) \cite{Aizenman:2019yuo}.} The standard physical picture, based on the Wilsonian renormalization group, is that such a limit should exist if the lattice model has a second-order phase transition that requires the tuning of the interaction coupling in order to be reached. In such a scenario, one would construct the continuum limit as a scaling limit in which the coupling is tuned to the phase transition in a coordinated way with the tuning of the lattice spacing to zero. Lacking an exact solution for the given model, such a scaling limit is a challenging task, especially when the universality class of the phase transition is described by an interacting theory. Luckily, even simple calculations and approximations can often help in establishing a correct physical picture. In particular, the Landau theory of critical phenomena provides very often a qualitatively (if not quantitatively) correct description of the phase transitions of a model.

In gravity, even the first step of formulating a lattice theory is nontrivial.\footnote{For the second step, that of the continuum limit, one hope would be to find a realization of Weinberg's asymptotic safety scenario \cite{Weinberg:1980gg,Niedermaier:2006wt,Bonanno:2020bil}.} Indeed, in general relativity the gravitational force is described in terms of the geometry of spacetime, whose dynamics is governed by background independent laws, so we should not rely on a fixed lattice background.
One proposal to construct a candidate lattice theory of Euclidean quantum gravity took shape in the 1980's in the context of noncritical bosonic string theory, and it is commonly known as dynamical triangulations (DT) \cite{Ambjorn:1997di}. It is based on Regge's coordinate-free approach to discrete gravity \cite{Regge:1961px}, in which continuous manifolds are replaced by piecewise flat manifolds (triangulations and their higher-dimensional generalizations, simplicial manifolds). However, unlike in Regge calculus, in DT one is prescribed to sum over all possible triangulations, with all edge-lengths fixed to the same value $a$, thus making DT also intrinsically background independent, in the sense that there is no fixed structure at all and no identifiable points. 
Originally DT was constructed for two-dimensional quantum gravity \cite{David:1984tx,Ambjorn:1985az,Kazakov:1985ea}, in which case it is related to matrix models \cite{DiFrancesco:1993nw}, and it had an exciting success in reproducing continuum results in the case of conformal charge $c<1$, thus proving that this way of discretizing the functional integral over geometries makes sense.\footnote{Even from a rigorous mathematical perspective, as reviewed in another chapter of the Handbook of Quantum Gravity \cite{Budd:2022zry}.} Later, it was generalized to three and four dimensions, where however the results were more disappointing: no second-order phase transition could be found \cite{deBakker:1996zx,Bialas:1996wu}. Moreover, the two phases of the model, separated by a first-order transition, were characterized by very degenerate geometries that did not approach a classical limit at large scales \cite{Ambjorn:1995dj}.\footnote{DT were  revived about a decade ago,  on the analytical side by the discovery of the large-$N$ limit of tensor models \cite{Bonzom:2011zz,Gurau:book}, prompting the hope that analytical control would show the way to improving DT, and on the numerical side by the exploration of some generalized models with hints of a new ("crinkled") phase \cite{Laiho:2011ya} (see also \cite{Benedetti:2011nn} for a meeting point of the two developments). Unfortunately, so far tensor models have not been able to solve the problem of escaping the branched polymer phase \cite{Gurau:2013cbh,Bonzom:2016dwy}, while the the crinckled phase of the generalized models turned out to be again rather unphysical, and still no second-order phase transition is in sight at finite coupling \cite{Ambjorn:2013eha,Coumbe:2014nea,Laiho:2016nlp}.}

\begin{figure}[htb]
\centering
\includegraphics[width=0.5\textwidth]{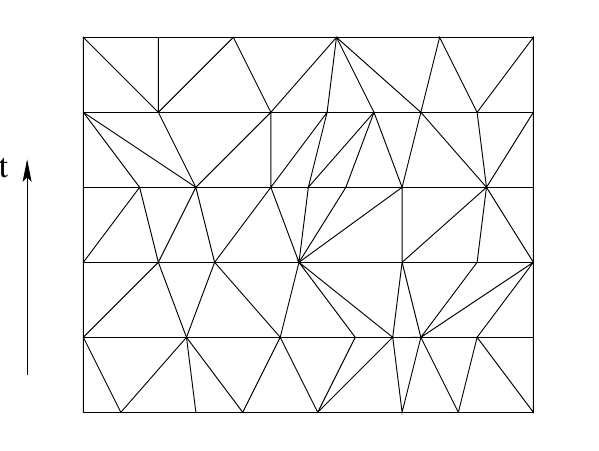}
\caption{\footnotesize Example of a triangulated portion of spacetime in the two-dimensional CDT model.
The flattened picture captures the way triangles are glued together, but does not represent
faithfully intrinsic distances, as edge lengths should be all equal.}
\label{fig:2dCDT}
\end{figure}

The state of affairs was significantly improved with the introduction of the model of causal dynamical triangulations (CDT) by Ambjorn and Loll  (initially in two dimensions \cite{Ambjorn:1998xu}, and later in three and four dimensions, together with Jurkiewicz \cite{Ambjorn:2001cv,Ambjorn:2005qt}). CDT modifies DT only by the requirement that the ensemble of triangulations be restricted with the requirement that each triangulation carries a non-singular foliation, i.e.\ a clear $(d+1)$-dimensional decomposition with $d$-dimensional slices (or leaves) of fixed topology, subsequent slices being always at fixed distance from each other (see \cite{Ambjorn:2001cv} for the precise construction, or the two-dimensional example in Fig.~\ref{fig:2dCDT} for an intuitive understanding).
Notice in particular that the foliation induces a natural notion of time as the direction perpendicular to the foliation, and a natural choice of time coordinate $t$, such that all the vertices on a given slice are at the same value of $t$.
With such a simple modification, originally motivated by the requirement of having always nonsingular light cones in a Lorentzian version of the model,\footnote{In this review we will only discuss models in Euclidean signature, as these are under better control, and in particular they are suitable for numerical simulations. The Lorentzian and Euclidean models can be easily Wick-rotated to each other at the discrete level. Given that Wick rotation is notoriously problematic in general relativity \cite{Visser:2017atf,Baldazzi:2018mtl}, the fact that in CDT (before any continuum limit) it poses instead no problems is a signal that perhaps the chosen ensemble of geometries is more restricted than one would have imagined.} the main problems of DT seemed to have been solved: a new phase of the model emerged with some classical behavior at large scales \cite{Ambjorn:2005qt}, separated from another phase by a second-order phase transition \cite{Ambjorn:2012ij} (see the reviews \cite{Ambjorn:2012jv,Loll:2019rdj} for a proper summary of results and a complete list references).

We should at this stage take a step back, and reflect upon the way phase diagrams are found in these models, which are effectively statistical models of random geometry. 
As for any statistical model, the standard way to chart a phase diagram is to identify suitable order parameters taking different values in the different phases. 
Such order parameters should necessarily correspond to expectation values of observables, meaning quantities which are independent of (unobservable) gauge choices.
Here one stumbles in the problem of observables in quantum gravity (e.g.\ \cite{Rovelli:1990ph,Giddings:2005id}): as in any diffeomorphism invariant and background independent approach to quantum gravity, observables in DT are necessarily nonlocal. Luckily, this is not an immediate problem for the phase diagram, as  order parameters are often nonlocal (e.g.\ the average magnetization in the Ising model), at least for the identification of homogenous phases, and indeed the two phases of DT can be distinguished by a global observable: the average linear size of the ``universe". In particular, studying the scaling of it with the total volume of the universe, one can define an effective Hausdorff dimension $d_H$ of the averaged universe. It turns out that for topological dimension $d=3$ and $4$, varying the lattice Newton constant, the effective dimension jumps from $d_H\sim 2$ (branched polymer phase) to $d_H\sim \infty$ (crumpled phase) \cite{Ambjorn:1995dj}.
Phases with the same Hausdorff dimensions were also found in CDT, but in addition a new phase was identified as well, with $d_H \sim d$  \cite{Ambjorn:2005qt}. 
Clearly, this was a great advance, providing a zeroth order test of classical behavior being possible in these models, but much more is needed in order to reconstruct the possible continuum limit.
For the latter, global observables are not very instructive, as for example it is hard to imagine how to construct and interpret an effective action for the Hausdorff dimension.

In CDT, thanks to the foliation, more observables are available, of a slightly more local nature: observables that are local in time but non-local in space.
The easiest observable of this type is the volume\footnote{In the two and three-dimensional cases, it would be more appropriate to call it  length and area, respectively, but for simplicity we stick to a general $d$-dimensional terminology.} of spatial slices $N_{d-1}(t)$, simply counting the number of $(d-1)$-simplices in the spatial slice at time $t$.
Armed with such observable, one can characterize the different phases of CDT in terms of volume profiles, and even derive an effective action describing the dynamics of the spatial volume \cite{Ambjorn:2004pw}. The latter will be our main focus in the following.

Given the premises above, a natural question arises: are the foliation and its associated observables compatible with a fully diffeomorphism invariant and background independent continuum limit? 
Originally, the foliation was thought only as a way of selecting less pathological geometries in a partially gauge-fixed (proper time gauge) version of discrete quantum gravity (see \cite{Dasgupta:2001ue} for a continuum point of view on this).
However, the existence of observables that are local in time raises two problems: first, they would be just a gauge artifact, and should not be true observables of the theory; second, a coarse-grained effective action depending on them would only be compatible with a gauge-fixed general-covariant action for very special values of the couplings.
The last point is extremely important, and, phrased differently, means that the most general and natural theory space in which a renormalization group flow of CDT should take place is that of theories invariant under foliation-preserving diffeomorphisms (FPD), and invariance under full (spacetime) diffeomorphisms would only be recovered on a submanifold of theory space.\footnote{We are being cavalier about the fact that any classical theory can be written in a fully diffeomorphism-invariant way. What we mean here is a theory that can be written in such a way without introducing other fields than the metric.}
FPD-invariant theories were introduced by Ho\v{r}ava \cite{Horava:2008ih,Horava:2009uw} in order to construct a gravitational theory that could be at the same time perturbatively renormalizable and unitary. This is achieved by introducing in the action higher derivatives in space, but not in time, which is possible thanks to the reduced diffeomorphism invariance. Such theories, of which several versions have been introduced and studied (see \cite{Wang:2017brl,Steinwachs:2020jkj} for reviews), are commonly known as Ho\v{r}ava-Lifshitz (HL) gravity, as they are the gravitational analog of theories with a Lifshitz critical point.
On the other hand, the space of FPD-invariant effective theories is more general and contains in particular also nonrenormalizable ones.

First possible hints of a relation between CDT and HL-type gravity theories were discussed in \cite{Horava:2009if,Benedetti:2009ge,Ambjorn:2010hu}, relying mostly on the presence of a foliation, on a qualitative resemblance between the phase diagram of CDT with the generic Lifshitz phase diagram, and on results concerning the spectral dimension (another useful notion of effective dimension). Other dynamical arguments have subsequently been added \cite{Budd:2011zm,Ambjorn:2013joa,Benedetti:2014dra,Benedetti:2016rwo}, and a longer list of arguments will be reviewed in Sec.~\ref{sec:conclusions}, but of course understanding the continuum limit of higher-dimensional background-independent statistical models like CDT is highly nontrivial, and therefore the precise relation is still open, in particular in four dimensions.

In any case, the fact that FPD-invariant effective theories are conceivable theories of dynamical geometry raises the following question: if CDT was describing a partially gauge-fixed but otherwise fully covariant theory, how would we construct a nonperturbative quantization of FPD-invariant theories in the spirit of dynamical triangulations, if not by introducing a foliation as in CDT? 
Most likely, the answer is that for FPD-invariant theories we should indeed use the same ensemble of triangulations as in CDT, in general with a different action, with more parameters (as for example in \cite{Anderson:2011bj}).
And if a fully diffeomorphism-invariant continuum limit could ever be reached from such generalized CDT models, this should be found within a subspace of their full theory space, i.e.\ by a careful fine tuning of the couplings.
Could we be so lucky that the standard CDT already sits in such a subspace? This might seem plausible, at least in two and three dimensions, where the CDT action is just a Regge discretization of the Einstein-Hilbert action, with cosmological and Newton constant.\footnote{Less so in four dimensions, where the CDT action has a third coupling, not appearing in
the Einstein-Hilbert action. However, this is a minor point with respect to what we want to stress here.}
Consider the example of a scalar model on a regular lattice. Typically, an anisotropic action will lead to an anisotropic continuum limit, while discrete rotation symmetry will result in full rotation invariance in the continuum limit (see for example \cite{Borji:2020snd}). That is, in this case, starting from an action which is a simple discretization of a rotation-invariant action, leads to the expected result, or in other words, the fine tuning is trivial.
But rotation invariance is a global symmetry. Things are more complicated for gauge symmetries, as their breaking leads to new degrees of freedom: for example, HL gravity has a new scalar degree of freedom besides the massless spin-2 graviton.
It is a well known fact that in the case that the gauge invariance of a quantum field theory is broken by a lattice or by a momentum cutoff, in order to restore it in the continuum limit, and thus recover the desired target theory, one needs a nontrivial fine tuning of all the possible breaking terms in the action, as for example extensively discussed in some lattice approach to chirality \cite{Testa:1997ia}, or in constructive \cite{Magnen:1992wv} and functional renormalization group \cite{Ellwanger:1994iz,Gies:2006wv} approaches to Yang-Mills theory.
It seems reasonable to expect that recovering full diffeomorphism invariance in CDT would require some fine tuning along those lines, although a luckier scenario (with no tuning of new couplings) cannot be ruled out as we still have a very limited understanding of the CDT measure and continuum limit in higher dimensions.

Let us recap the key elements of the discussion above. DT models provide a diffeomorphism invariant and background independent discretization of a quantum theory of dynamical geometry, but unfortunately they do not have a second order phase transition and they are dominated by very degenerate geometries. In order to solve these problems, in CDT one reduces the ensemble of triangulations by introducing a built-in regular foliation. 
However, the foliation being a sort of background structure, this likely breaks part of the diffeomorphism invariance of the target theory (general relativity), and in order to restore that, one would presumably need to fine tune the model, as well as add new counterterms to the action. The latter part of this summary is so far conjectural, because, due to the complexity of the models, which are mostly studied by Monte Carlo simulations, we cannot rule out in all certainty that the standard CDT action would be a lucky choice leading to a full diffeomorphism invariant theory in the continuum.

We will argue below that in two dimensions it is quite clear that indeed in the continuum limit CDT is described by an FPD-invariant theory, that in three dimensions there is some evidence supporting the same conclusion, and that even in four dimensions there are some indications in the same direction. To that end, we will focus on the effective dynamics of the volume of the spatial slices, which can be seen as an application of Landau theory to CDT.
In particular, we will argue that by a deeper study of the effective action governing such observable, one can in principle distinguish between a gauge-fixed general-covariant action and an FPD-invariant one.

The rest of the review, mainly based on the results of \cite{Benedetti:2014dra,Benedetti:2016rwo}, is organized as follows. In section \ref{sec:motivations}, we review the main ideas of the Landau theory of critical phenomena, the challenges of applying it to a background independent model, and how instead a Landau free energy is straightforwardly obtained for two-dimensional CDT, in a direct top-down approach. Armed with that, one recognizes in the continuum limit the two-dimensional version of HL gravity.
In section \ref{sec:3d-review}, we review the setup and numerical results for the dynamics of the spatial volumes in three-dimensional CDT.
In section \ref{sec:BIB}, we review a bottom-up proposal (based on numerical data and on the intuitions we tried to motivate in this introduction) for the Landau free energy associated to such spatial volumes, viewed as (time-dependent) order parameter of the model. We also review the numerical simulations of the coarse-grained model based on such Landau free energy, showing that qualitative features of the CDT phase diagram are reproduced.
An extremization analysis of the continuum limit of the Landau free energy is reviewed in section \ref{sec:analysis}, providing in particular an analytical understanding of the condensation phenomenon observed in the semiclassical phase of CDT, and one more indication that the correct effective action of the model sits in the space of FPD-invariant theories.
Lastly, we conclude in section \ref{sec:conclusions} with a short summary, and a discussion of our current understanding of the relation between CDT and HL-like gravity theories.

%%%%%%%%%%%%%%%%%%%%%%%%%%
\section{Landau Theory and CDT}
\label{sec:motivations}
%%%%%%%%%%%%%%%%%%%%%%%%%%

The goal of Landau theory (e.g.\ \cite{Goldenfeld:1992qy}) is to provide a theoretical framework for understanding and predicting phase transitions in statistical systems.
In this respect, it has represented an important landmark in statistical physics, successfully explaining many phenomena. It is also well-known to have strong limitations, predicting wrong scaling exponents below critical dimensions, and this led to the development of the renormalization group. However, in general Landau theory remains a valid tool for a first analysis, at least at the qualitative level, for example for understanding the structure of the phase diagram of a given model.

The central aspect of Landau theory is that it postulates that phase diagrams can be explained in terms of order parameters, and that these are effectively governed by a coarse-grained free energy functional, known as Landau free energy, which depends parametrically on the coupling constants of the system. The minima of the Landau free energy determine the thermodynamically favored configuration of the order parameters, and different favored configurations correspond to different phases. Typically, there will be a \emph{disordered phase}, with vanishing order parameters and unbroken symmetries, and one or more \emph{ordered phases}, with non-vanishing order parameters breaking some or all of the symmetries of the system. Therefore, given a statistical model, it is crucial to identify the order parameters and construct the corresponding Landau free energy.

There are two main approaches for constructing the Landau free energy: a phenomenological (bottom-up) approach, writing it as an expansion in the order parameters, constrained by the symmetries, as in effective field theories; and a top-down approach, deriving the Landau free energy by explicitly coarse-graining the microscopic model. The latter is of course seldomly applicable in practice, but it provides a very instructive way of thinking about the Landau theory, and it is hence worth of being briefly recalled (we essentially follow \cite{Goldenfeld:1992qy} for this).

Consider a spin system with Hamiltonian $H(\{s\})$, where $\{s\} = \{ s_i \mid i\in \cL \}$ is a configuration of spin variables $s_i$ on the $N$ sites of the lattice $\cL$.
The partition function is
\be
Z_{\rm spin} =  \sum_{\{s\}} e^{-\b H(\{s\})} \, .
\ee
Next, partition the lattice into blocks $\cB_r$, labeled by $r$, which can be thought of as a site of the coarse grained lattice $\cL'$. Each block contains $N(r)$ sites, such that $\sum_r N(r) = N$.
We define the Landau free energy $L(\{m\})$, where $\{m\} = \{ m(r) \mid r\in \cL' \}$, as the Gibbs free energy for the system, constrained to be in a configuration compatible with a local magnetisation configuration specified by $\{m\}$, i.e.:
\be \label{eq:LandauF}
e^{-L(\{m\})} =  \sum_{\{ s_i \mid i\in \cL,\, \f{1}{N(r)}\sum_{j\in\cB_r} s_j =  m(r) \}} e^{-\b H(\{s\})} \, .
\ee
From the definitions, it follows straightforwardly that the partition function can be expressed in terms of the Landau free energy as
\be \label{eq:LandauZ}
Z_{\rm spin} = \sum_{\{m\}} e^{-L(\{m\})} \, .
\ee
It is clear that in this construction, $L(\{m\})$ can also be thought as a Wilsonian effective Hamiltonian. This is in particular the case because the local magnetisation $\{m\}$ is a similar (albeit coarse grained) type of variable as the original spin variables $\{s\} $, in the sense that we went from some spin variables to some other spin variables (perhaps with a different set of values, due to the averaging).
 The idea is however more general, and depending on the model and the relevant order parameter, one can define different coarse-graining procedures and end up with very different variables.

\

\noindent {\bf Landau theory and quantum gravity.}\quad
Applying a similar construction to quantum gravity is of course marred with difficulties, in particular concerning the identification of order parameters, as observables are notoriously problematic in quantum gravity. Diffeomorphism invariance and background independence constrain observables to be non-local quantities, which are not very suitable for reconstructing a local effective action. Moreover, coarse graining presents several practical and conceptual obstacles, due to the absence of a background structure.
Nevertheless, inventive approaches to the problem have been devised, with coarse graining procedures introduced for constructing a renormalization group flow in DT \cite{Johnston:1994qi,Thorleifsson:1995ki,Ambjorn:1996hu,Renken:1996kf,Henson:2009fy} and in spin foams \cite{Markopoulou:2002ja,Oeckl:2002ia,Bahr:2012qj,Dittrich:2013uqe,Steinhaus:2020lgb}.
More to the point of interest here, mean field models of DT have also been proposed \cite{Renken:1997na,Bialas:1998ci}.
The latter involved no coarse graining, they were rather in the spirit of defining approximated Hamiltonians (or actions, in the parlance of Euclidean QG) much like what is done when replacing the Ising model Hamiltonian with one in which spin interacts not just with its neighbour, but with an average of all the other spins (the mean field). Again this procedure, which is straightforward in the Ising case, is not at all obvious in DT, and in fact the mean field models for DT were postulated, rather than derived from the model itself.

Once more, in CDT things are (slightly) different thanks to the foliation. Small-step blocking is still a very clumsy procedure, but we do have a natural partitioning of the lattice into large blocks: the leaves of the foliations. We can therefore define a Landau free energy by summing over all triangulations that on each slice lead to a prescribed value for a nonlocal \emph{slice observable}. 
As anticipated, the choice of slice observable on which we will concentrate is the volume of the slice, which is essentially the only one studied so far, but it would be interesting to explore other options.
As a way of clarifying this set of ideas, and of introducing the more difficult higher-dimensional cases, we start by analyzing the two-dimensional CDT model in this perspective.

%%%%%%%%%%%%%%%%%%%%%%%%%%
\subsection{CDT in a nutshell}
%%%%%%%%%%%%%%%%%%%%%%%%%%

The CDT approach exactly parallels the usual lattice field theory with one fundamental difference: the fixed lattice is replaced by an ensemble of random triangulations.
This is required by the fact that gravity is a theory of dynamical geometry, with no background spacetime fixed a priori.

More concretely, one defines an ensemble of ``triangulations" to work with, a triangulation being defined by a simplicial manifold,  i.e.\ a collection of $d$-dimensional flat simplices (the generalization of triangles and tetrahedra) glued along their $(d-1)$-dimensional faces and such that the neighbourhood of any vertex is homeomorphic to a $d$-dimensional ball.  A dynamical triangulation is one in which all the simplices are taken to be equilateral.  In the simulations we usually work with dynamical triangulations having a fixed number of $d$-simplices $N$, which we will denote $\cT_{N}$. The ensemble of such triangulations $\{\cT_{N}\}$ is obtained by gluing the $N$ simplices in all possible ways allowed by the simplicial manifold condition, and respecting a chosen topology.
Furthermore, to avoid the sick behaviour that was found in the old models of dynamical triangulations, CDT models have one further restriction on the ensemble: only triangulations with a global time foliation, with respect to which no spatial topology change occurs, are allowed. For more details on the geometrical meaning of this restriction and on its implementation see \cite{Ambjorn:2001cv}.

Once the ensemble is specified one can construct the partition function (Euclidean version of the path integral) as
\be \label{Z}
Z = \sum_N \sum_{\cT_{N}} \tfrac{1}{C(\cT_{N})}\, e^{-S(\cT_{N})} \, ,
\ee
where $S(\cT_{N})$ is the bare action, and $C(\cT_{N})$ is the order of the automorphism group of $\cT_{N}$, a symmetry factor naturally appearing when summing over unlabeled triangulations (e.g.\ \cite{Ambjorn:2012jv}). Since we wish to recover general relativity in the classical limit, it is customary to use as a bare action the Einstein-Hilbert action adapted to a simplicial manifold, which is known as the Regge action. On a dynamical triangulation, the Regge action takes the very simple form
\be \label{action}
S(\cT_{N}) = \kappa_d N - \kappa_{d-2} N_{d-2} \, ,
\ee
where $\kappa_d$ and $\kappa_{d-2}$ are two coupling constants depending on the cosmological and Newton's constant appearing in the Regge action, and $N_{d-2}$ is the number of $(d-2)$-dimensional subsimplices (also called bones or hinges).

In principle one could use a different action, with more parameters, but at this stage this would only complicate the analysis of the results, and in a minimalist attitude such a generalization of the CDT models is usually postponed till the moment (if ever) at which the model itself will ask for such an extension. For example in $3+1$ dimensions a new parameter $\D$ has been introduced in the action, without which no physically interesting region would exist in the phase diagram \cite{Ambjorn:2005qt}.
Furthermore, we need to remember that, for fixed topology, only $d/2$ (for $d$ even) or $(d+1)/2$ (for $d$ odd) among the values $\{N_0,N_1,...N_{d-1},N_d=N\}$ are independent. Hence, for $d=3$ and $d=4$ only two of such variables are independent, and as a consequence, if we want to keep the action linear in $N_j$, we only have two coupling constants. The counting changes in CDT because we can distinguish subsimplices whose vertices are all on one slice from those having vertices on two adjacent slices, and additional variables can be introduced in order to keep track of that.
However, new topological relations are found too. The counting for CDT was carried out in \cite{Ambjorn:2001cv}, and one has that for $d=4$ there are 10 variables and 7 constraints, leaving 3 independent variables, a fact that was used in  \cite{Ambjorn:2005qt} to introduce the new parameter. In $d=3$ the situation is instead unchanged with respect to the DT case, as there are 5 constraints for 7 variables, and hence again only 2 independent variables. For this reason it is not possible to introduce in three dimensions the analogue of the new parameter used in four dimensions.

 Typically, in CDT one fixes also the number of time slices $T$, which in fact can be seen as a new free variable, appearing thanks to the foliation.
We therefore often deal with the canonical partition function 
\be \label{Z_CDT}
Z(N,T) =  \sum_{\cT_{N,T}} \tfrac{1}{C(\cT_{N,T})}\, e^{-S(\cT_{N,T})} \, ,
\ee
where $\cT_{N,T}$ now stands for a foliated $d$-dimensional simplicial manifold with a fixed number $N$ of $d$-simplices and a fixed number $T$ of spatial slices, plus some boundary conditions to be specified. Typically one considers spatial slices with closed topology. In the time direction the boundary conditions are either periodic ($T+1\sim 1$) or open (slices at time 1 and $T$ as boundaries). The latter case can be viewed as a propagator from an initial to a final geometry in $T-1$ steps.

Notice that thanks to the foliation we can also introduce a transfer matrix $W$, given by the one-step propagator for the spatial slices, such that the grand canonical partition function for CDT with periodic boundary conditions in the time direction is $Z(T) =\Tr[W^{T}]$ (see for example \cite{Ambjorn:2001br}).

We are now ready to define more precisely the Landau free energy $L(\{m\})$ for spatial volume observable. In analogy with the definition \eqref{eq:LandauF} for the spin system, with the slices playing the role of the blocks that partition the lattice, we define
\be \label{eq:Landau-CDT-can}
e^{-L(\{m\}; N,T)}  =  \sum_{\cT_{N,T}} \tfrac{1}{C(\cT_{N,T})}\, e^{-S(\cT_{N,T})} \prod_{i=1}^T \d_{m_i,N_{d-1}(i)}\, ,
\ee
where $N_{d-1}(i)$ denotes the number of $(d-1)$-simplices at slice $i$ in the simplicial manifold $\cT_{N,T}$.
Similarly, we can define a grand canonical Landau free energy $L(\{m\}; T)$ by summing over simplicial manifolds $\cT_T$ with unconstrained number of $d$-simplices.

Assuming that, at least in some approximation, $L(\{m\}; T)$ only contains nearest-neighbour interactions among the spatial volume variables, 
i.e.\ $L(\{m\}; T)\approx \sum_{i=1}^T \tilde{L}(m_i,m_{i+1})$, then the grand canonical partition function will admit also a transfer matrix formulation with a transfer matrix $\tilde{W}=e^{-\tilde{L}}$ for just the spatial volume variables:\
\be \label{eq:Landau-CDT-gcan}
Z(T) = \sum_{\{m\}} e^{-L(\{m\}; T)}  \approx \Tr[\tilde{W}^T] \,.
\ee
It has been conjectured in \cite{Ambjorn:2001br} that, for $d>2$, summing the geometry-to-geometry transfer matrix $W$ over the boundary geometries, with fixed volumes taken to be large, a volume-to-volume transfer matrix is obtained, which then we could identify with  $\tilde{W}$. 
Such a conjecture seems closely related to the short-range assumption about $L(\{m\}; T)$, although it is probably stronger, in the sense that the former implies the latter, but not the other way round.

%%%%%%%%%%%%%%%%%%%%%%%%%%
\subsection{Balls-in-boxes models}
\label{Sec:BIB}
%%%%%%%%%%%%%%%%%%%%%%%%%%

Before continuing our discussion of CDT, it is useful to introduce the so-called balls-in-boxes (BIB) models, which will play the role of effective models of CDT.

The statistical models known as  balls-in-boxes (or zero-range process in the nonequilibrium version) are an interesting and versatile class of models which have been extensively studied in the statistical mechanics literature (see for example \cite{Evans-review}).
They have also been used as mean field models of DT \cite{Bialas:1996eh,Bialas:1997qs,Bialas:1998ci}, and
later, also as effective models for the spatial volume dynamics of CDT in 3+1 dimensions \cite{Bogacz:2012sa}. We briefly review in this section their definition and some of their relevant properties.

A  BIB model is defined as a one-dimensional lattice with $T$ sites (boxes) to each of which is associated an integer number $m_i\geq m_{\rm min}\geq 0$ (the number of balls in box $i\in\{1,2, ..., T\}$). The total number of balls is fixed to be $M$.
The canonical partition function of the statistical model is given as
\be\label{Z_BIB}
\begin{split}
Z_{\rm BIB}(T,M) &= \sum_{m_1=m_{\rm min}}^M ...  \sum_{m_T=m_{\rm min}}^M \d_{M,\sum_i m_i } \prod_{i=1}^T g(m_i,m_{i+1}) \\
 &= \sum_{\{m_j\} } e^{-S(\{m_j\})}  \d_{M,\sum_i m_i } \, ,
 \end{split}
\ee
with $m_{T+1}=m_1$.   The last expression highlights the interpretation of such models as discretised one-dimensional path integrals with action $S(\{m_j\})$. However, the fact that the configurations are subjected to the nonlocal constraint
\be \label{constraintM}
\sum_{i=1}^T m_i  = M \, ,
\ee
%.
sets such models apart from standard (Euclidean) path integrals, thus justifying the use of a specific name such as balls-in-boxes.  
The weight function $g(m,n)$ (or the action $S(\{m_j\})$) defines the particular model. In the standard BIB models there is no nearest-neighbour interaction, meaning that $g(m_i,m_{i+1}) =g(m_i)$; the model above is a generalisation studied in \cite{Evans-prl,Waclaw:2009zz}.

Often, in particular for an analytical approach, it is useful to work in the grand canonical ensemble, for which the partition function reads
\be \label{Z_BIB-gc}
\begin{split}
Z_{\rm BIB-gc}(T,z) &= \sum_M Z_{\rm BIB}(T,M)\, z^M \\
   &= \Tr [ \hat W^T ]\, ,
\end{split}
\ee
where we introduced the transfer matrix
\be
\hat W_{m,n} = z^{(m+n)/2} g(m,n) \, .
\ee
In the light of this relation, $g(m,n)$ is sometimes referred to as {\it reduced transfer matrix}.

One interesting feature of these models is that they can exhibit a condensation phenomenon. In the original BIB models, with $g(m_i,m_{i+1}) =g(m_i)$, this means that for certain values of the parameters the model enters into a phase dominated by configurations completely localised at one (random) site. 
The mechanism behind such condensation remains similar in the more general models, but the nearest-neighbour interaction allows the condensate to spread over a region whose width scales with a power of the total volume $M$.
As we will see, it is this type of condensation which provides the basis for an explanation of the droplet configuration in CDT based on the much simpler BIB models.

It should also be self-evident from the formulas above that the BIB models are straightforwardly interpretable as effective models of CDT in the Landau coarse grained picture of \eqref{eq:Landau-CDT-can}, in particular under the assumptions leading to \eqref{eq:Landau-CDT-gcan}.
However, in principle we do not need to conjecture any relation between the BIB transfer matrix and the full transfer matrix of CDT: the first equality in  \eqref{eq:Landau-CDT-gcan} is an exact rewriting, while for the second we only need an assumption of short-range interaction between the slices.

%%%%%%%%%%%%%%%%%%%%%%%%%%
\subsection{Landau theory of two-dimensional CDT}
\label{sec:2dCDT}
%%%%%%%%%%%%%%%%%%%%%%%%%%

In two dimensions, only one of the $N_k$ variables is independent, e.g.\ the number of triangles $N_2$. This is clear also from a continuum perspective, as  the curvature term is topological, hence at fixed topology the Einstein-Hilbert action reduces to a simple volume term. The two-dimensional CDT model \cite{Ambjorn:1998xu} at fixed topology has thus only one coupling, the bare cosmological constant, while Newton's constant plays no role. The corresponding partition function has therefore the interpretation of generating function for the number of triangulations of the type in figure \ref{fig:2dCDT} with fixed number of triangles $N$ and some given boundary conditions.

The two-dimensional model of CDT can be solved exactly by various means \cite{Ambjorn:1998xu,DiFrancesco:1999em,DiFrancesco:2000nn,DiFrancesco:2001xur}, so it would seem pointless to look for a Landau theory. 
However, the latter appears very naturally in the process of solving the model, and it is therefore a useful example of top-down approach to a Landau free energy in a quantum gravity model.
In this case, the partition function for two-dimensional causal triangulations with fixed $N$ and $T$ (denoted ($\cT_{N,T}$) can be written as
\be \label{eq:2dBIB}
\begin{split}
Z_{\rm 2d-CDT}(N,T) &= \sum_{\cT_{N,T}} 1 \\
&= \sum_{l_1=1}^{N} ...  \sum_{l_T=1}^{N} \d_{N_2,2\sum_i l_i } \prod_{i=1}^T g(l_i,l_{i+1}) 
\\
&\equiv \sum_{\{l\}} e^{-L(\{l\};N,T)} \, ,
 \end{split}
\ee
with $l_{T+1}=l_1$ (periodic boundary conditions in the time direction), and
\be
g(l_i,l_{i+1}) =  \f{ (l_i + l_{i+1} )! }{ l_i! \, l_{i+1}! } \, ,
\ee
counting the number of triangulations of a strip with boundary lengths $l_i$ and $l_{i+1}$, with open boundary conditions in the spacelike direction.\footnote{It basically counts the number of ways we can place $l_{i+1}$ balls in $l_i+1$ boxes, so that in this case the reduced transfer matrix is itself a BIB model (with trivial reduced transfer matrix).}

In the second line of \eqref{eq:2dBIB} we recognize a BIB model (compare to \eqref{Z_BIB}), while in the last line we defined the function $L(\{l\};N,T)$, which we interpret as a Landau free energy. Indeed, comparing to \eqref{eq:LandauF} and \eqref{eq:LandauZ}, we see that in the definition of $L(\{l\};N,T)$ we are summing over triangulations with a block constraint, the blocks here being the spatial slices of each triangulation, and the constraint fixing the number of edges in each block.\footnote{Such a nonlocal observable on a codimension-one slice is reminiscent of similar quantities considered in anisotropic systems, e.g. the Ising model with anisotropic competing next-to-nearest-neighbor interactions \cite{Murtazaev:2013}.}

Notice that the Landau free energy $L(\{l\};N,T)$ contains a global constraint, but we can get rid of it by switching to the grandcanonical ensemble.
That is, by summing over $N$ with a Boltzmann weight $e^{-\kappa_2 N}$, we obtain:
\be
L_{\rm gc}(\{l\};T,\kappa_2) = 2\kappa_2 \sum_i l_i - \sum_i \ln\left(g(l_i,l_{i+1})\right) \,.
\ee
It should be stressed that the second term is a purely entropic contribution, with no lattice coupling constant.

\

\noindent {\bf Continuum limit.}\quad 
The model is exactly solvable \cite{Ambjorn:1998xu,DiFrancesco:1999em,DiFrancesco:2000nn,DiFrancesco:2001xur}, and a continuum limit can be taken on the solution. It is however instructive to see what we can learn by first taking a continuum approximation, and then minimizing the continuum Landau free energy.
Using Stirling's formula one finds that, for large $l_i$ and $l_{i+1}$, and small  $(l_{i+1}-l_i)/(l_i + l_{i+1})$,
\be \label{2d-transfer}
g(l_i,l_{i+1}) \sim 2^{l_i + l_{i+1} } e^{-\f{(l_{i+1}-l_i)^2}{l_i + l_{i+1}}} \, .
\ee

It is easy to take  the continuum limit of the exponent in \eqref{2d-transfer}, thus obtaining
\be \label{S_2d}
S_{\rm 2d-HL}[\ell] = \int_{-\f{\t}{2}}^{\f{\t}{2}} dt\,
\f{\dot{\ell}(t)^2}{4 \ell(t)}
 \, .
\ee
Such action\footnote{Here and in the following we will generally refer to the continuum limit of the Landau free energy as an (effective) action.} can be interpreted as an action for the length $\ell(t)$ of the slices, describing two-dimensional HL gravity in proper-time gauge, as noticed in \cite{Ambjorn:2013joa} (see also the dedicated chapter of the Handbook of Quantum Gravity \cite{Sato:2022ory}).

It is well known (see for example \cite{Mattei:2005cm} and references therein) that in a path integral quantisation of gravity in the proper-time gauge we loose the Hamiltonian constraint, unless we integrate over the total proper time. Such integration is not performed in CDT when computing finite-time propagators (by the definition of such observables), or when doing simulations with periodic boundary conditions in time,  for obvious practical reasons.
However, it should be stressed that in principle this only means that the integral over time is postponed to a later stage.  Indeed, in 1+1 dimensions this integral can be carried out exactly, leading to a solution of the quantum Hamiltonian constraint, or Wheeler-DeWitt equation \cite{Ambjorn:1998xu}.
What we want to stress here is that in order to make contact between continuum models and CDT results with fixed total time, one should not try to impose the Hamiltonian constraint in the former.
Bearing this in mind, we can try to see what a semiclassical analysis of \eqref{S_2d} tells us. 

Solving the equations of motion of \eqref{S_2d} with the constraint $V_2=\int_{-\f{\t}{2}}^{\f{\t}{2}} dt \ell(t)$, and with periodic boundary conditions in time, one finds either constant or oscillating solutions.
The constant solution is in fact unique, due to the volume constraint: $\ell(t)=V_2/\t$. The oscillating solutions form a discrete set, due to the periodicity condition. 
Plugging these solutions into the action \eqref{S_2d}, since $\ell(t)\geq 0$ and $\dot{\ell}(t)\neq 0$ (except at isolated points) for the oscillating solution, it is obvious that the constant solution has the least action (it evaluates to zero) and therefore it must dominate the path integral. 
Note that the Hamiltonian constraint would fix the amplitude independently of the total time $\t$, and therefore we would have periodic solutions only for special values of $\t$.
We could consider also droplet configurations of the kind that we will introduce in higher dimensions, but the constant solution would still dominate over them, as the action \eqref{S_2d} is non-negative, and equal to zero if and only if $\dot{\ell}(t)= 0$ for all $t$.

The dominance of the constant solution is in complete agreement with the Monte Carlo snapshots from numerical simulations \cite{Ambjorn:1999gi}, which, unlike the higher-dimensional models, show no sign of translational symmetry breaking, and average to a constant profile. 
This can also be seen from the analytical solution \cite{Ambjorn:1998xu}, which gives a constant $\la \ell(t) \ra \propto 1/\sqrt{\L}$, where $\L$ is the continuum cosmological constant in the grand canonical ensemble.\footnote{One might worry that in the case of periodic boundary conditions a simple average will always lead to a translational-invariant result. The usual way to deal to avoid blurring a broken phase with an average over vacua is to introduce an explicit symmetry breaking term in the model and check if the symmetry breaking persists when this is continuously removed.
For example, one could break explicitly the translational symmetry by introducing initial and final boundaries of fixed length, and check that for large $\t$ the bulk is approximately constant, but this would be a long and unnecessary parentheses here. As we will discuss later, in numerical simulations there is a more practical way to effectively select a single vacuum.}
In any case, one should also remember that in $1+1$ dimensions $\sqrt{\la \ell^2 \ra-\la \ell \ra^2} \propto 1/\sqrt{\L}$, i.e.\ fluctuations have the same magnitude as the average configuration, thus hiding any possible classical behaviour of the latter.

The role of higher order terms in the Stirling approximation has been studied in \cite{Ambjorn:2015gea}, where it was shown that including enough matter fields the leading order correction (a logarithmic potential term in the exponent of \eqref{2d-transfer}) becomes important and leads to a phase transition to a droplet phase, similar to the one we will describe below.

\

\noindent {\bf Effective theory vs minisuperspace models.}\quad 
We should stress once more that \eqref{S_2d} is not a minisuperspace approximation for the quantization of the bare theory. In fact, the bare action in two-dimensional CDT is trivial.
Equation \eqref{S_2d} is instead obtained nonperturbatively as a purely entropic contribution from the sum over all triangulations.
This also shows as a proof of principle that even if we start with a bare action of general relativity, the structure of triangulations of our ensemble can determine a more general universality class. Of course such a phenomenon should not come as a surprise: while it is easy to define a diffeomorphism-invariant action (in the continuum), it is highly nontrivial to define a regularized diffeomorphism-invariant measure for the path integral, and therefore it is precisely there that a possible breaking of gauge symmetries might occur.

%%%%%%%%%%%%%%%%%%%%%%%%%%
\section{Spatial volume dynamics in three-dimensional CDT}
\label{sec:3d-review}
%%%%%%%%%%%%%%%%%%%%%%%%%%

In this section we concentrate on the case $d=3$, for which very few analytical results are known (e.g.\ \cite{Ambjorn:2001br,Benedetti:2007pp,Durhuus:2014dbl}) because of the difficulty in solving statistical models in dimensions higher than two. Therefore, most of the current understanding of this case derives from Monte Carlo simulations.

In the simulations one typically uses the topological constraints in order to trade the variable $N_1$ for $N_0$, which is easier to keep track of, and thus replace \eqref{action} by
\be \label{action-3d}
S(\cT_{N}) = \kappa_3 N - \kappa_0 N_0 \, .
\ee
Furthermore, as we mentioned, in the computer simulations we work at fixed volume, and hence we replace \eqref{Z} by the partition function for the canonical ensemble,
\be \label{Z_N}
Z_N =  \sum_{\cT_{N}} \tfrac{1}{C(\cT_{N})}\,e^{\kappa_0 N_0} \, ,
\ee
where we have made use of the simple form of the action \eqref{action-3d}. Note that the grand canonical partition function $Z$ is the discrete Laplace transform of $Z_N$ with respect to $N$.

The expectation value of an observable $\cO$ is calculated as
\be
\la \cO \ra_N = \frac{1}{Z_N} \sum_{\cT_{N}} \tfrac{1}{C(\cT_{N})}\,  e^{\kappa_0 N_0} \cO(\cT_{N})\, ,
\ee
which is related to the grand canonical expectation value as a function of $\kappa_3$ via
\be
\la \cO \ra = \frac{1}{Z} \sum_N e^{-\kappa_3 N} Z_N \la \cO \ra_N \, .
\ee

\

\noindent {\bf Simulations.}\quad 
Simulations were performed in \cite{Benedetti:2009ge,Benedetti:2014dra} using the Markov-chain Monte Carlo technique.  An adaptation of some previously existing code for the Monte Carlo simulations (used in \cite{Ambjorn:2000dja}) was used for this purpose.  The code generates a finite set of sample configurations $\{\cT^{(1)},...,\cT^{(M)}\}$ according to the probability distribution $\cP(\cT) = \f{1}{Z} e^{-S(\cT)}$.  One then approximates the expectation value of an observable by its arithmetic mean across these samples:
\be
\la \cO \ra \approx \f{1}{M} \sum_{j=1}^M \cO(\cT^{(j)}).
\ee

As usual in three-dimensional CDT simulations, the spacetime topology was fixed to $S^2 \times S^1$, i.e.\ spherical spatial sections and cyclical time.\footnote{Simulations with different boundary conditions in the time or spatial directions have been performed in \cite{Cooperman:2013mma} and \cite{Budd:2013waa}, respectively, and the results are consistent with those reviewed here.} Values of $N$ up to a maximum of 200k$\,\equiv 2 \times 10^5$ were studied, although some errors for the larger values of $N$ are greater since less configurations could be generated to be averaged over, within practical time constraints.  All simulations were carried out with coupling constant $\kappa_0=5$, in the phase where previous evidence points to the emergence of a well-behaved (three-dimensional) geometry.  The total number of time-steps was set to $T=96$.

%-------------------------------------------------------------
\subsection{The volume data}
\label{Sec:data}
%-------------------------------------------------------------

The observable we study here is the volume of the spatial slices, which in the three-dimensional CDT model corresponds to the number of spatial triangles $N_2(i)$ as a function of discrete time $i$. 
Because the triangulation is connected we always have $N_2(i)>0$. Furthermore, because we restrict to simplicial manifolds, the smallest triangulation of a two-sphere has four triangles, giving
\be \label{minN2}
N_2(i) \geq n_{\rm min} =4\, .
\ee

\begin{figure}[ht]
\centering
\includegraphics[width=.9\textwidth]{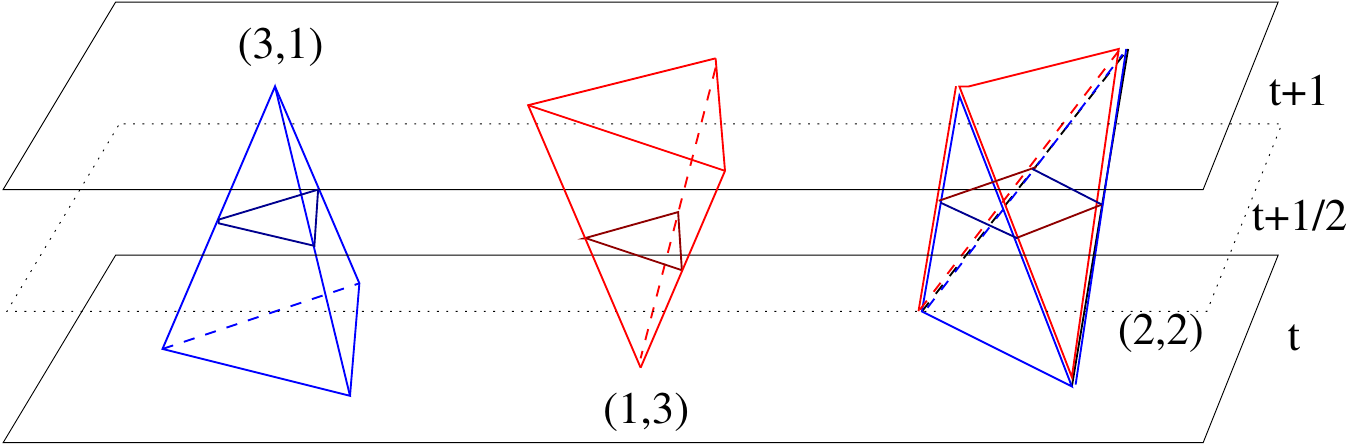} %100 percent
\caption{\footnotesize The three types of tetrahedral building blocks in three-dimensional CDT, and their
intersections with the $t+1/2$ plane.}
\label{fig:blocks}
\end{figure}

Another possible observable is the volume of the spatial slices at half-integer values of time, which amounts to a weighted sum of the number of (3,1) and  of (2,2) tetrahedra (see Fig.~\ref{fig:blocks}) between slices $i$ and $i+1$. We expect that in the phase of extended geometry (where $N_{(3,1)}\sim 2 N_{(2,2)}$) any such differences in definitions of volume as a function of time should be irrelevant in the continuum limit.
%%%%%%%%%%%%%%%%
%%%%%%%%%%%%%%%%

\begin{figure}[ht]
\centering 
\includegraphics[width=\textwidth]{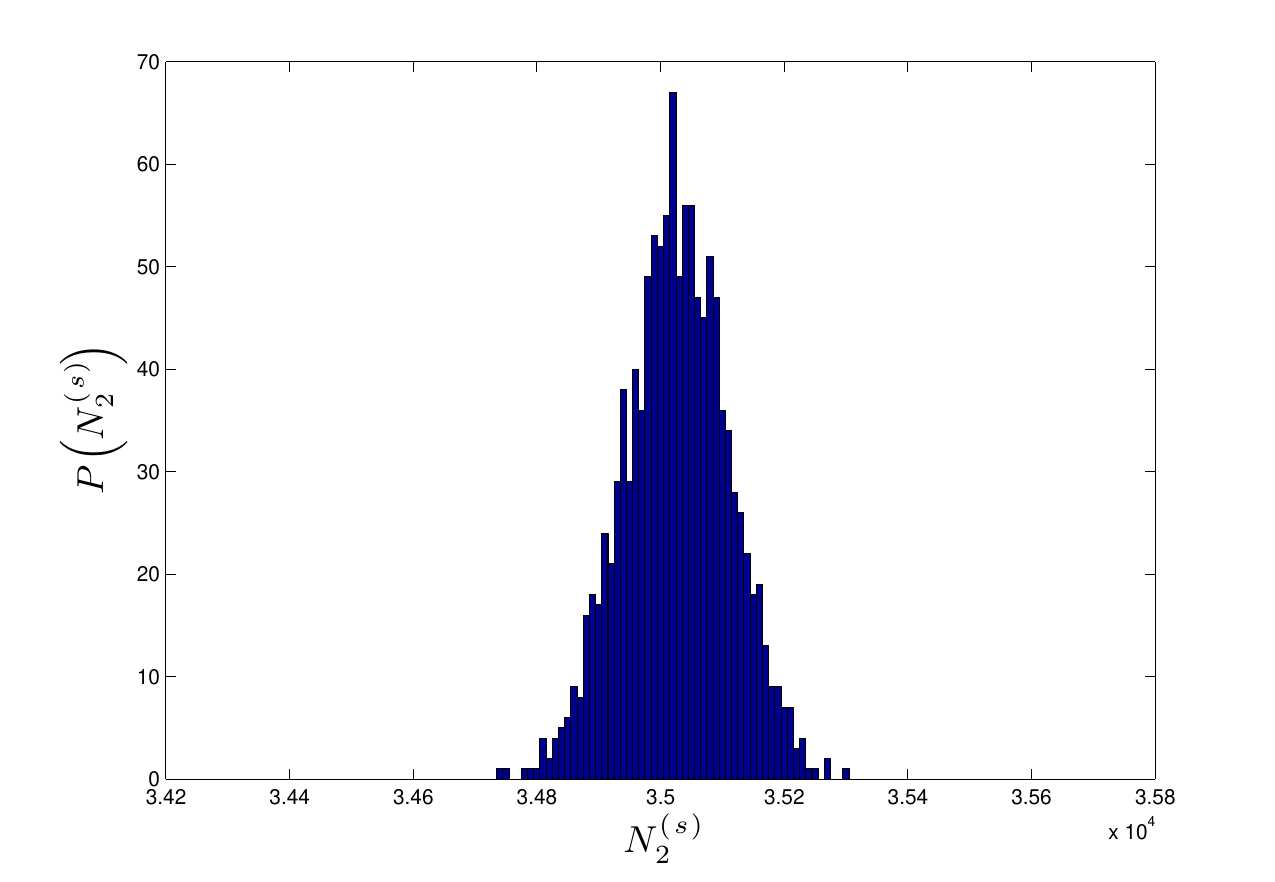} 
\caption{\small{The distribution of the total number of spatial triangles $N_2^{(s)}$ (grouped in bins of size 10), for $N_3$=100k. The expectation value is $\la N_2^{(s)}\ra =35026$, with standard deviation $\s=85$.}}
\label{Fig:N_2-tot}
\end{figure}

Note that a triangle is always shared by two tetrahedra, so that, for the number $N_2^{(s)}$ of spatial triangles, we have
\be \label{N2s}
N_2^{(s)} = \sum_{i=1}^T N_2(i) = \f12 N_{(3,1)}\, ,
\ee
which in the extended phase we expect to be roughly one third of the total volume $N_3$.
For the value of $\kappa_0$ we used,  the distribution of $N_2^{(s)}$ is very peaked (for $N_3=$100k the relative standard deviation is only $0.2\%$, see Fig.~\ref{Fig:N_2-tot}), and we find that $N_{(2,2)}$ is just slightly smaller than a third of the total volume.
%%%%%%%%%%%%
%%%%%%%%%%%%

\begin{figure}[ht]
\centering 
\includegraphics[width=\textwidth]{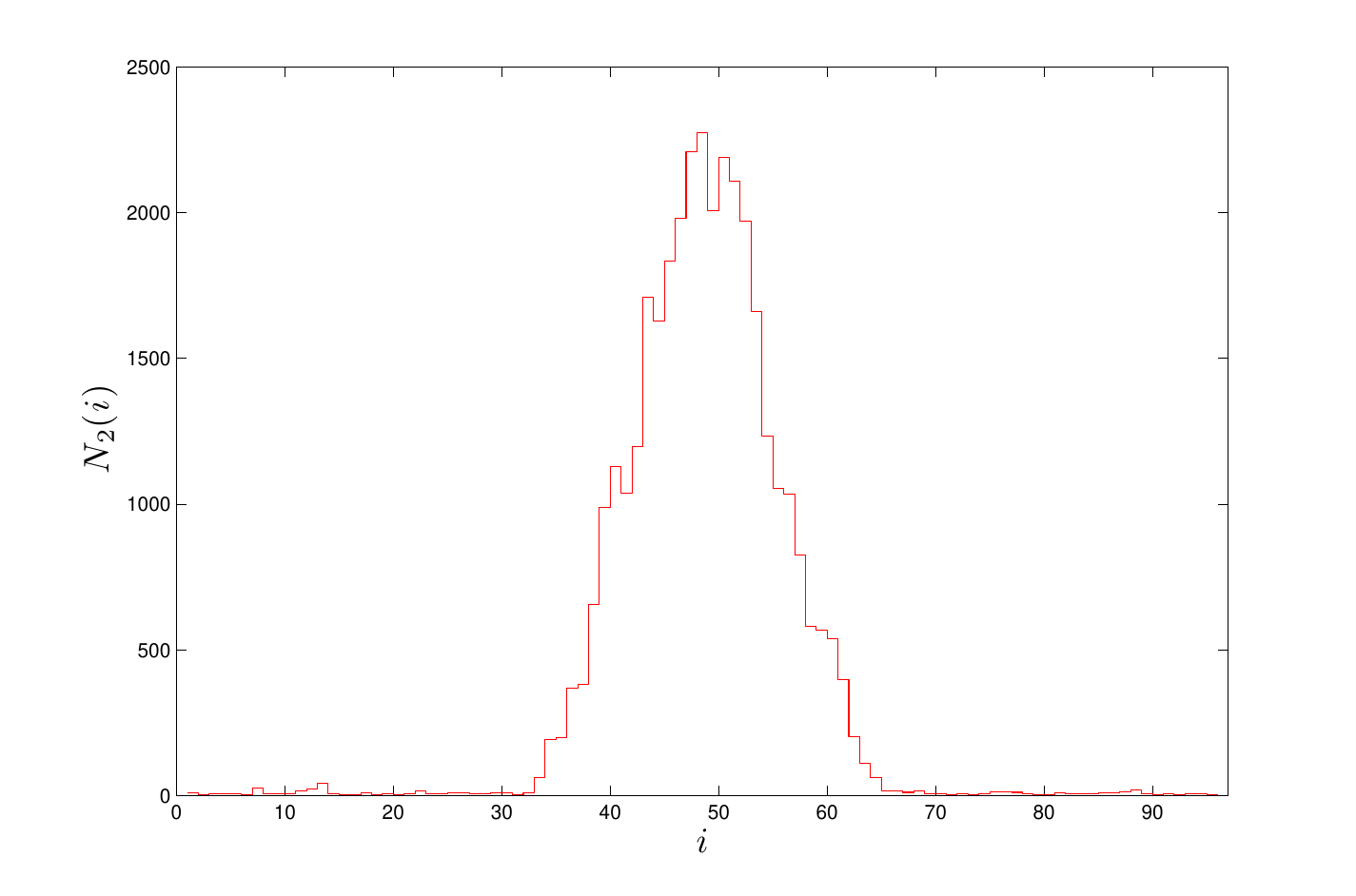} 
\caption{\small{The spatial area $N_2(i)$ as a function of time, from a single configuration taken at random in the data set of Monte Carlo simulations for $N_3$=100k. 
}}
\label{Fig:hist-N_2}
\end{figure}

\begin{figure}[ht]
\centering 
\includegraphics[width=\textwidth]{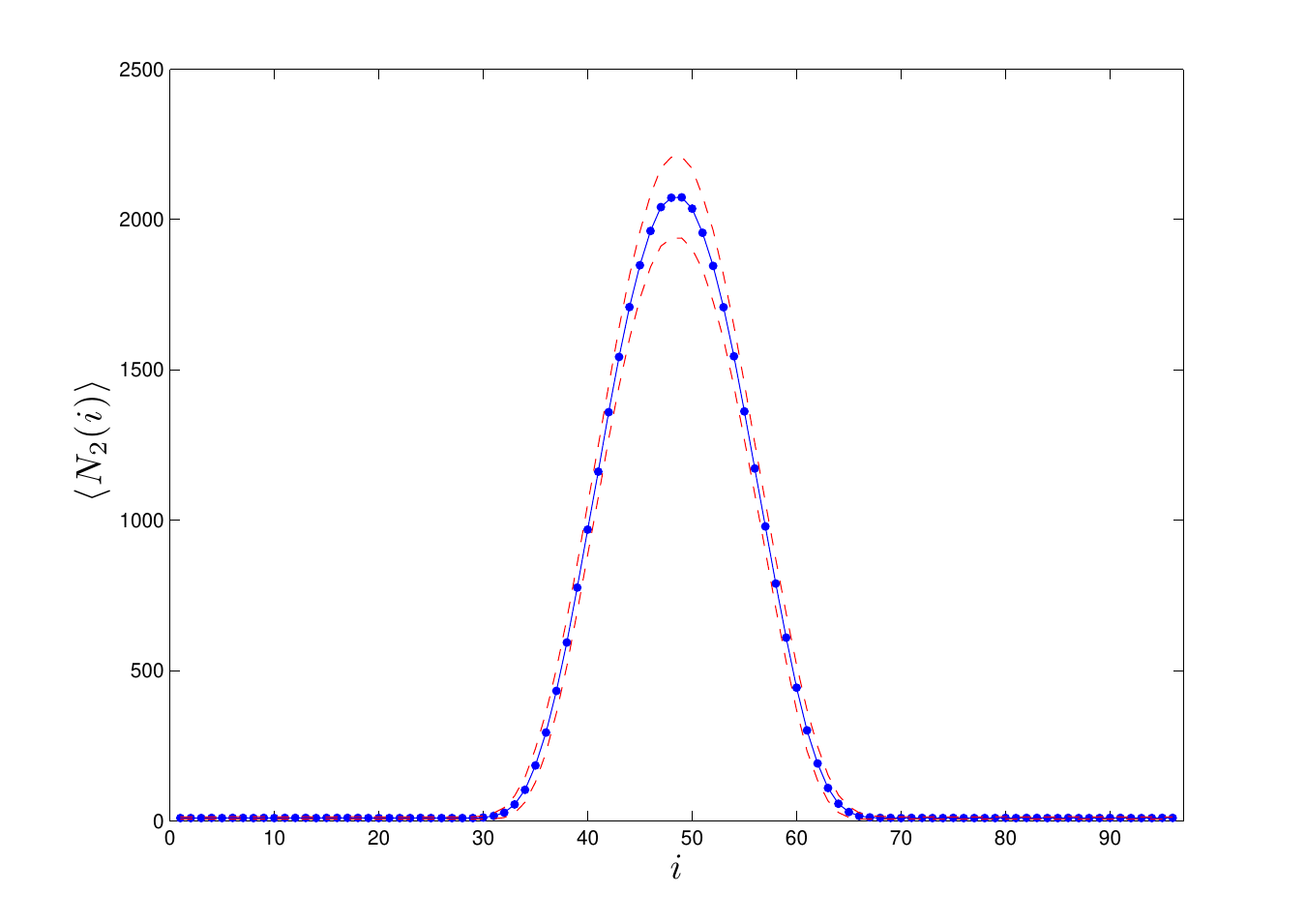} 
\caption{\small{The mean spatial area as a function of discrete time, for $N_3$=100k. The blue dots represent the mean area $\la N_2(i)\ra$  of the volume data at time $i$, while the blue line is just an interpolation curve. Error bars (estimated for example by the jackknife method) are roughly of the size of the dot radii, or smaller, and thus we chose to omit them for clarity (this is done throughout the paper).
The dashed red lines are interpolating curves for $\la N_2(i)\ra \pm \s/2$, where $\s=\sqrt{\la (N_2(i))^2\ra-\la N_2(i)\ra^2}$ is the size of the fluctuations.}}
\label{Fig:N_2}
\end{figure}

In  Fig.~\ref{Fig:hist-N_2}  we show the volume profile $N_2(i)$ from a snapshot of a Monte Carlo simulation.
One notices immediately  a phenomenon of spontaneous (translational) symmetry breaking: the Monte Carlo configuration shows a condensation of the volume around a specific time. Averaging over Monte Carlo configurations, the translational symmetry gets restored, but this is because a blind average amounts to summing over the degenerate vacua of the broken phase. As usual, in order to see spontaneous symmetry breaking we should select only one vacuum. For example, in a spin system this is done by first coupling the spins to an external magnetic field, which is then switched off. In the CDT simulations, we select one vacuum by performing a recentering of the Monte Carlo configurations.
In practice, we have to find the center of volume $t_{CV}(j)$ for each Monte Carlo configuration $j$, and we have to shift time so that $t'_{CV}(j)=T/2$ for every configuration in the new time variable. We performed this operation following the method given in \cite{Ambjorn:2008wc,Gorlich:2011ga}. 
Once the data are centered in this way, it makes sense to study the average of  $N_2(i)$. A plot of $\la N_2(i)\ra$,  together with fluctuations, is displayed in Fig.~\ref{Fig:N_2}.

%%%%%%%%%%%%
%%%%%%%%%%%%
%%%%%%%%%%%%
%%%%%%%%%%%%

The volume profile has a characteristic extended part (typically referred to as the \textit{blob} or \textit{droplet}), and a long flat tail (referred to as the \textit{stalk}).  Within the latter, the spatial volume is very close to its kinematical minimum, $\f1m \sum_{i=1\ldots m}^{i\in {\rm stalk}}\la N_2(i)\ra \equiv n_s \sim 10$, and it is independent of the total volume. Most of the total volume is therefore concentrated in the blob.
We will discuss in the following sections how to explain this condensation phenomenon, and which function best describes the volume profile.

%-------------------------------------------------------------
\subsection{The continuum limit}
\label{s:c_limit}
%-------------------------------------------------------------

We conclude this section by discussing how the continuum limit $a\to 0$ is investigated on the basis of the simulation data.
All observables and couplings in the simulations are given as dimensionless numbers, as such are also all the quantities appearing in \eqref{action-3d} and the definition of the model. Length dimensions are introduced by multiplying the quantity of interest by the appropriate power of the cutoff $a$, the length of spacelike edges.\footnote{Despite the Euclidean signature we use the words spacelike and timelike to distinguish the orientation of edges with respect to the foliation.}
For example, we can write $\t = a \, \a\,   T$ for the time interval, where we have introduced a parameter $\a>0$ to allow a different length of the timelike edges. Next, we can write $V_2 = \f{\sqrt{3}}{4} a^2 N_2$ for the volume of a slice (the numerical prefactor being the area of an equilateral triangle of unit side),  $V_3 = v_{(3,1)}(\a) a^3 N_{(3,1)} + v_{(2,2)}(\a) a^3 N_{(2,2)}$ for the total volume, etc. Here, $v_{(3,1)}(\a)$ and $v_{(2,2)}(\a)$ stand for the volume of the $(3,1)$ and $(2,2)$ simplices with spatial edges of length one, and time edges of length $\a$ (they both coincide with the equilateral tetrahedron for $\a=1$, with volume $v_3=1/6\sqrt{2}$, see  \cite{Ambjorn:2012jv}).

An essential part of the continuum limit procedure is to have $N\to\infty$ and $a\to 0$ in such a way that the physical volume
\be \label{volume}
V \sim a^d N
\ee
remains finite. 
This implies that when we multiply a dimensionless quantity by $a^n$, in order to switch to its dimensional counterpart, in practice we multiply it by $N^{-n/d}$.
For example, in CDT we also demand that $\t  \sim N^{-1/d}T$ stays finite.

In the thermodynamic limit we expect that, besides the total volume \eqref{volume}, also other dimensional large-scale observables will become independent of the cutoff $a$.
Therefore, we expect to see scaling behavior when working with simulations at sufficiently large $N$, i.e.\ we expect that an observable $\la \cO(i) \ra_N$, depending on the discrete time variable $i=1,\ldots,T$, will satisfy for some fixed (continuum time) $t$:
\be \label{fss}
N^{-n/d} \la \cO(N^{1/d} t) \ra_N = N'^{-n/d} \la \cO(N'^{1/d} t) \ra_{N'} \,,
\ee
where the value of the observable at non-integer time is to be intended from an interpolating fit.
The natural expectation would be that \eqref{volume} and \eqref{fss} hold with $n$ given by the expected dimension of the observable, but this is not guaranteed a priori and it is instead used as a check of the good classical properties of the model, as we do in the following.
Moreover, not all quantities will show scaling: the dimensionless version of some quantity might remain independent of $N$. The typical example is the correlation length $l_c$, from $\la \cO(i) \cO(j) \ra_N-\la \cO(i) \ra_N\la\cO(j) \ra_N \sim e^{-|i-j|/l_c}$ (at large $|i-j|$); in order to obtain a finite dimensionful correlation length $\ell_c = a l_c\sim N^{-1/d} l_c$, one needs to tune the couplings towards a critical point where $l_c$ diverges.
We will not be concerned with correlation lengths here, but we will see another example of quantity that stays at the cutoff scale, i.e.\ the volume of the stalk in the droplet phase.

%%%%%%%%%%%%
\begin{figure}[ht]
\centering 
\includegraphics[width=\textwidth]{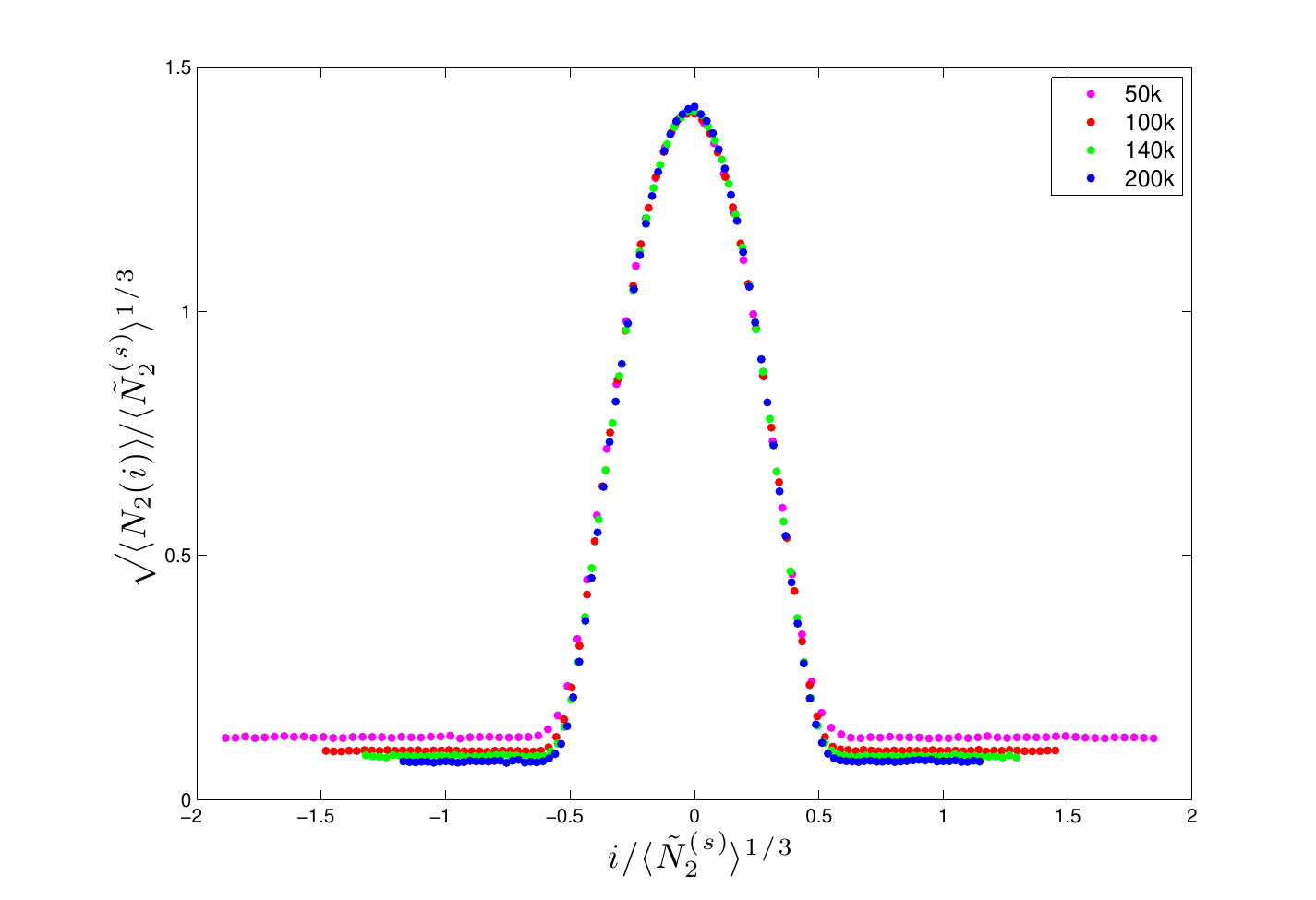} 
\caption{\small{The square root of the mean spatial area as a function of discrete time (shifted so that the peak is at the origin). Both the mean area and the time variable are rescaled with the appropriate power of the average total volume $\tilde N_2^{(s)} = N_2^{(s)} - n_s T$, in order to display scaling.}}
\label{Fig:scaling}
\end{figure}
%%%%%%%%%%%%
Fig.~\ref{Fig:scaling} shows a scaling of the type \eqref{fss}. Here we have plotted  $\la N_2(i)\ra^{1/2}/ \langle \tilde N_2^{(s)}\rangle^{1/3}$ as a function of $i/ \langle \tilde N_2^{(s)}\rangle^{1/3}$, for different data sets corresponding to different total volumes $N$. 
Following the procedure used in \cite{Ambjorn:2005qt}, we used   
%%
%\be \label{red-vol}
$\la \tilde N_2^{(s)}\ra = \la N_2^{(s)} \ra - n_s T$,
%\ee
%
instead of $\la N_2^{(s)}\ra$ (or $N_3$, which as we saw, is proportional to it) in the rescaling because we know that the volume in the stalk does not scale.
The plot clearly shows that the superposition is extremely good inside the droplet, while in the stalk the rescaled volume goes to zero for growing $N_3$.

Another dynamical quantity that goes to zero in the continuum is the size of the fluctuations around the average, which we expect  to be controlled by Newton's constant.
Indeed, without any fine tuning of the dimensionless couplings, all the couplings having positive (negative) length dimension will go to zero (infinity) as $a\to 0$. As Newton's constant has the dimension of length for $d=3$, it is expected to go to zero in the naive continuum limit (in four spacetime dimensions, where Newton's constant has dimension of length to the second power, this has been observed in \cite{Ambjorn:2008wc}). 
A finite coupling in the continuum could perhaps be attained by fine tuning the bare coupling to a second order phase transition.

%%%%%%%%%%%%%%%%%%%%%%%%%%
\section{Effective models}
\label{sec:BIB}
%%%%%%%%%%%%%%%%%%%%%%%%%%

In this section we review how in four and (in more detail) three dimensions the dynamics of spatial volumes in CDT can be effectively described by BIB models. Minimization of the BIB model action, in accordance with its interpretation as a Landau free energy, leads to a good qualitative description of the full CDT phase diagram.

%-------------------------------------------------------------
\subsection{Four-dimensional CDT}
%-------------------------------------------------------------

In  \cite{Bogacz:2012sa}, Bogacz et al.\ studied a BIB model which is roughly a discretised version of a minisuperspace model corresponding to four-dimensional general relativity, and which was originally considered in \cite{Ambjorn:2004pw} to explain the data from the simulations.
It was found that this very simple model can account for many of the observed features of CDT in four dimensions, including its rich phase diagram. In particular, a droplet phase was found, which has remarkable similarities to the extended phase of CDT, although of course the comparison is limited to the behaviour of the spatial volume against time.

The model is defined by the following reduced transfer matrix,
\be \label{BIB4d}
g(m_i,m_{i+1}) = \exp\left( -c_1 \f{2 (m_{i+1}-m_i)^2}{m_i + m_{i+1}} - c_2 \f{m_i^{1/3} + m_{i+1}^{1/3}}{2} \right)\, .
\ee
In the continuum, this corresponds to the following action for the volume  $V_3(t)$ of the 3-dimensional spatial slices,
\be \label{S_mini4d}
S_{\rm 4d-mini}[V_3] = \f{1}{2 G} \int_{-\f{\t}{2}}^{\f{\t}{2}} dt \, \left( c_1 \f{\dot V_3^2(t)}{ V_3(t)} + c_2 V_3^{1/3}(t) \right) \, ,
\ee
which is precisely of the type obtained from a minisuperspace reduction of general relativity (where $c_1=1/\cN$, and $c_2=9 (2\pi^2)^{2/3} \cN$, $\cN$ being the lapse function).
This is the action that was conjectured from the very beginning by Ambjorn \textit{et al.}~as an effective description for the extended part of the universe (or blob) in their simulations \cite{Ambjorn:2004pw}, a conjecture which was further corroborated over the years \cite{Ambjorn:2005qt,Ambjorn:2007jv,Ambjorn:2008wc,Ambjorn:2011ph,Ambjorn:2012pp}.
The novelty in \cite{Bogacz:2012sa} was the suggestion that the same effective action can explain much more than just the dynamics inside the blob.
Effectively, we can reinterpret the insight from \cite{Bogacz:2012sa} as the observation that the BIB model with reduced transfer matrix \eqref{BIB4d} provides a good guess for a Landau free energy, i.e.\ in a bottom-up approach.

Notice that in the minisuperspace action that one would derive from the Einstein-Hilbert action the constant $c_1$ would be negative, while for a meaningful use of \eqref{S_mini4d} as an effective action, and from comparison to the numerical data \cite{Ambjorn:2004pw}, we must take it to be positive.\footnote{The choice looks of course irrelevant if all we have is this action, as we can just change its overall sign, but this is not so if in the full Einstein-Hilbert action we want also the spin-two modes to have the good sign.} 
It would thus seem that in CDT the conformal factor problem \cite{Gibbons:1978ac} affecting Euclidean quantum gravity is solved, perhaps cured by nonperturbative contributions from the functional measure, as suggested in \cite{Dasgupta:2001ue}, effectively recovering the rotated minisuperspace model of \cite{Hartle:1983ai}. We will comment further on this below, arguing that in fact a simpler explanation is possible.

Another important difference between the usual minisuperspace model of general relativity and the BIB model, is that in the latter there is no analogue of the lapse to be integrated in the partition function and neither there is a summation/integration over $T$. As a consequence, there is no Hamiltonian constraint to be imposed in the semiclassical analysis. 
Above, we emphasised that the same situation should be expected in CDT, where the distance between one spatial slice and the next is constant and the total time extension of the universe is fixed in all simulations to date.
This is also supported by the strong evidence from numerical simulations that the BIB model is a good effective description for CDT.

The equations of motion derived by varying \eqref{S_mini4d} with respect to $V_3(t)$ are
\be \label{eom4d}
c_1\left( \left( \f{\dot V_3}{V_3} \right)^2 -2 \f{\ddot V_3}{V_3}  \right)  + \f{c_2}{3} \f{1}{V_3^{2/3}} - \L =0 \, ,
\ee
where the cosmological constant $\L$ is introduced as a Lagrange multiplier, to be fixed by imposing the volume constraint. If we were to impose also the Hamiltonian constraint $H\equiv c_1 \dot V_3^2(t)/V_3(t) - c_2 V_3^{1/3}(t) +\L V_3(t) = 0$, the combined system of equations would reduce to a first order differential equation (by deriving the Hamiltonian constraint with respect to time, and eliminating $\ddot V_3(t)$ between the two equations).
As such, its solutions would have only one free integration constant, which could be fixed for example by demanding that the maximum of $V_3(t)$ be at $t=0$.
The solution would then be the ``$\cos^3$'' solution discussed by Ambjorn et al.  in \cite{Ambjorn:2007jv,Ambjorn:2008wc}.
However, without the Hamiltonian constraint the equation remains second-order, and thus there is one more free parameter, which allows us to perform a minimisation of the on-shell action. 

For such a minimization one does not need to restrict to class $C^2(S^1)$ functions, since for a well defined action \eqref{S_mini4d} it is sufficient that $V_3(t)\in C^1(S^1)$. This allows the authors of \cite{Bogacz:2012sa} to consider droplet configurations that are not classical solutions, but that are expected on the basis of simulations and heuristic arguments.\footnote{One should bear in mind that this sort of analysis is not aimed at reproducing the detailed profile of CDT at the junction between blob and stalk. In the junction region of the droplet we expect the effect of subleading terms in the action to be non-negligible. Furthermore, the time-interval mesh in the CDT simulations is not fine enough to reveal much about the smoothness of the average profile at such junction.}
Using $V_4 = \int_{-\f{\t}{2}}^{\f{\t}{2}} dt V_3(t)$ for the total volume in the continuum, they obtain the following expression for the dominant contribution to the path integral in a particular region of the the phase diagram:\footnote{As we will see in more detail for the three-dimensional case, a number of parametric conditions must be satisfied for such configuration to dominate.}
\be \label{fullsol4d}
\bar V_3(t) = \begin{cases} \f{3 \om V_4}{4} \cos^3(\om t) \, ,  &  \text{for } t \in [-\f{\pi}{2\om},+\f{\pi}{2\om}]\, ,  \\  0  \, ,  &  \text{for } t \in [-\f{\t}{2},-\f{\pi}{2\om})\cup (+\f{\pi}{2\om},+\f{\t}{2}]\, ,  \end{cases}
\ee
where 
\be \label{frequency4d}
\om = \f{\sqrt{2} }{3 V_4^{1/4} } \left( \f{c_2}{c_1} \right)^{3/8}\, .
\ee
Notice that $V_3(t)=0$ obviously minimises the action \eqref{S_mini4d} for positive $c_1$ and $c_2$. However, alone it would fail to satisfy the volume constraint, while a balance between the zero and the  ``$\cos^3$'' solutions wins the energy balance, resulting in a condensation. Interestingly, the blob part of \eqref{fullsol4d}, together with \eqref{frequency4d}, corresponds to the solution obtained by imposing also the Hamiltonian constraint, but this seems a mere coincidence. 

The crucial point to be made here is that the presence of a potential term in \eqref{S_mini4d} allows non-constant configurations such as \eqref{fullsol4d} to dominate the path integral.  Thus, the conclusions derived in this case are qualitatively different from those derived from \eqref{S_2d} in the two-dimensional case.

The result is very interesting because it shows how the reduced model in four dimensions not only reproduces the extended part of the universe, but also its stalk, and it gives a prediction for their relative time extension.
The same model also predicts other phases \cite{Bogacz:2012sa} that resemble the so-called phases A and B of CDT \cite{Ambjorn:2005qt}.
However, it should be stressed that it fails to capture the more recently discovered "bifurcation" phase of CDT  \cite{Ambjorn:2014mra,Ambjorn:2016mnn}, for which the spatial volume is not sufficient and another order parameter is required.

It is also worth noticing that from the Einstein-Hilbert action the ratio $c_2/c_1$ would be fixed, thus leaving us with no parameters for a fit to the CDT simulations. Since the width of the universe (for fixed $V_4$) depends on the bare coupling $\kappa_0$ \cite{Ambjorn:2008wc}, this is another indication that the effective action describing the volume dynamics is different form what would be expected from general relativity.
In other words, the $(\Delta,\kappa_0)$ space of CDT coupling constants translates into a $(c_1,c_2)$ plane of effective coupling constants in the BIB model, while we would expect only one in the GR minisuperspace model.
On the other hand, such argument is not conclusive, as the additional parameter could turn out to be irrelevant (in the renormalization group sense) in the continuum limit, and thus amount just to an innocuous lattice artefact.

We will now argue that in three dimensions a stronger case can be made for a departure of the effective BIB model from a GR minisuperspace model.

%-------------------------------------------------------------
\subsection{Three-dimensional CDT}
\label{Sec:3d_BIB}
%-------------------------------------------------------------

In three dimensions the minisuperspace reduction of the Einstein-Hilbert action contains no potential term, i.e.\ the action for the spatial areas
 in the proper-time gauge is exactly of the same form as \eqref{S_2d}, but with the length $L(t)$ replaced by the area of the spatial slices $V_2(t)$. As in the two-dimensional case, in the presence of a volume constraint  $V_3 = \int dt V_2(t)$ one finds oscillating solutions of the equations of motion, with a volume profile  $V_2(t)\sim \cos^2 (\om t)$ which is compatible with the one observed inside the blob of Fig.~\ref{Fig:N_2}. For this reason, such action has been suggested as an effective action for CDT \cite{Ambjorn:2000dja,Ambjorn:2002nu} (see also \cite{Cooperman:2013mma}).
However, as pointed out  in \cite{Benedetti:2014dra}, in the light of the results on BIB models for two and four-dimensional CDT, the similarity of the general relativistic minisuperspace action in three dimensions to \eqref{S_2d} immediately raises the question of how such an effective action could ever explain the important differences between two and three-dimensional CDT. Given that the two-dimensional case reduces exactly to a BIB model, and the four-dimensional case is well described (in its spatial volume dynamics) by a BIB model, we would expect a similar model to give also a good approximation for the three-dimensional model. 

Of course an important difference between two and three dimensions is in the scaling of dimensionful quantities. Most importantly, Newton's constant is dimensionless for $d=2$ while it has dimension of length for $d=3$. As a consequence, in the (naive, not fine-tuned) continuum limit, Newton's constant (and with it the fluctuations around the average volume of the slices) scales to zero in the latter case, while it stays constant in the former. In two dimensions, as there are no other scales besides the cosmological one, the size of the fluctuations in the continuum limit is as large as the expectation value, thus blurring any classical behaviour. On the contrary, in three dimensions the fluctuations go to zero and the classical (mean field) behaviour should dominate.

Regardless of how the fluctuations behave, it turns out that the action derived from the Einstein-Hilbert action fails in reproducing the CDT results in an important way.
Repeating the analysis of Bogacz et al.\ for three dimensions we simply have to set $c_2=0$ in the previous subsection, or just recall what we have said about the two-dimensional case.
From \eqref{frequency4d} we then find that $\om= 0$, and the width of the droplet diverges.
However, \eqref{frequency4d} does not hold in this case, as it would violate the condition $\pi/\om<\t$ which is to be assumed in \eqref{fullsol4d}.
On the other hand we have already explained what happens in the two-dimensional case, and going to three dimensions we simply have to replace  $V_2\to V_3$ and $\ell(t)\to V_2(t)$.
The droplet solution which minimises the action is obtained for $\om=\pi/\t$. However, it is easily checked that in such a case the action is strictly positive, while for $V_2(t)=V_3/\t$ the action vanishes. We conclude that a BIB model derived from the three-dimensional Einstein-Hilbert action would predict a constant average profile for the two-dimensional volumes.
This is also supported by the numerical simulation of   \cite{Bogacz:2012sa}, as for $c_2=0$ and $c_1>0$ the model defined by \eqref{BIB4d} lies in the correlated fluid phase, not the droplet phase.

We are left with the challenge of explaining the droplet condensation of three-dimensional CDT as a condensation of a BIB model.  This also provides us with an extraordinary opportunity to test corrections to the effective action of \cite{Ambjorn:2000dja,Ambjorn:2002nu}. In higher dimensions, such corrections are expected to be subdominant with respect to the the linear spatial curvature term coming from the Einstein-Hilbert action (the one multiplied by $c_2$ in \eqref{S_mini4d}). 
Fortunately, however, the $d=3$ case is an exception to this, because this linear curvature term results in an irrelevant constant term for the potential, and hence drops out of the story, making higher order corrections relevant.

In view of that, and inspired by the reasonings outlined in the introduction, as well as by the fact that there is presently no evidence for the presence of higher-order time derivatives in the effective action \cite{Ambjorn:2011ph}, in \cite{Benedetti:2014dra} we proposed an alternative effective theory for the spatial areas, one derived from HL gravity.

\

\textbf{Starting point:} One sets off with the most generic, projectable, $z = 2$ action for Ho\v rava-Lifshitz gravity in three dimensions \cite{Horava:2008ih}:
\begin{equation}
  \label{eq:hl-action}
  S_{\rm 3d-HL} = \frac{1}{16\pi G} \int \extd t\,\extd^2x\; \cN\sqrt{g}
                  \left\{
                      %\sigma
                      ( K_{ij}K^{ij} - \lambda K^2)
                      + 2\Lambda
                      - bR
                      - \gamma R^2
                  \right\}\,.
\end{equation}
As discussed in the introduction, HL\ gravity describes a class of metric theories supporting a preferred foliation. Thus, the space-time diffeomorphism symmetry is broken down to foliation-preserving diffeomorphisms.\footnote{Such foliation-preserving diffeomorphisms are described by time-reparameterizations and spatial diffeomorphisms of the form:
  \begin{equation}
    \label{eq:hl-diffeos}
    t \rightarrow t + \xi^0(t)\qquad\textrm{and}\qquad x^i \rightarrow x^i + \xi^i(t,x)\,,
  \end{equation}
  where $(t,x^i)$ are co-ordinates in an atlas of charts adapted to the foliation.
}

In \eqref{eq:hl-action}, the action is presented in terms of ADM variables: $\cN$ is the lapse function, $g$ is the determinant of the spatial metric, $R$ is its Ricci scalar, $K_{ij}$ is the extrinsic curvature associated to the leaves of the foliation, while $K$ is its trace.

The action contains the familiar parameter pair $(G, \Lambda)$, corresponding to Newton's constant and the cosmological constant, respectively. The parameter $\sigma = \pm1$ neatly encapsulates some metric signature information.  Meanwhile $\lambda$, $b$ and $\gamma$ characterize the deviation from full diffeomorphism invariance. Indeed, for $\lambda = b = 1$ and $\gamma = 0$, one recovers the Einstein-Hilbert Lagrangian in Euclidean 
signature: $2\Lambda-\mathcal{R}$, where $\mathcal{R}$ is the spacetime Ricci tensor.\footnote{Notice that we have changed the overall sign of the action with respect to \cite{Benedetti:2014dra,Benedetti:2016rwo}, because in retrospective we find the present choice more clear. Ultimately the two ways of presenting this motivational steps are equivalent as they lead to the same effective action (thanks also to the freedom we have in choosing $\lambda$ in HL gravity).}

The theory is said to be \textit{projectable} if one imposes at the outset that the lapse function is spatially constant: $\cN = \cN(t)$.

The $z$-\textit{exponent} refers to half the maximal order of spatial derivatives appearing in the action. Thus for $z = 2$, it may contain at most four spatial derivatives, permitting the inclusion of the $R^2$ term.

We consider spacetimes with topology $S^{2}\times S^1$, that is, spherical spatial slices and a compactified time. To implement this, we impose periodic boundary conditions in time, with period $\tau$.

\textbf{Mini-superspace reduction:} We perform a mini-superspace reduction to constant lapse, vanishing shift vector (hidden so far in the extrinsic curvature) and spatial metric $g_{ij} = \phi^2(t) \hat{g}_{ij}$, where $\hat{g}_{ij}$ is the standard metric on the unit sphere.  
And in order to compare to CDT in the canonical ensemble, we replace the cosmological term as:
\be
\frac{2\L}{16\pi G} \int \extd t\,\extd^2x\; \cN\sqrt{g} \to \frac{2\L}{16\pi G} \left(V_3 - \int \extd t\,\extd^2x\; \cN\sqrt{g} \right)\,,
\ee
and treat $\L$ as a Lagrange multiplier.

With a redefinition of parameters,\footnote{We define:
  \begin{equation}
    \label{eq:hl-mini-parameters}
    \kappa^2 =  \frac{NG}{1-2\lambda},\qquad
    \omega^2 = \frac{N^2\Lambda}{1-2\lambda}\,,\qquad
    b' = -\frac{\cN^2b}{1-2\lambda}\,,\qquad
    \xi = \frac{2\cN^2\gamma}{1-2\lambda}\,.
  \end{equation}
}
the action becomes:
\begin{equation}
  \label{eq:hl-mini-action}
  S_{\textrm{3d-HL-mini}}[\phi] = \frac{1}{2\kappa^2}\int_{-\frac{\tau}{2}}^{\frac{\tau}{2}} \extd t \left\{\dot{\phi}^2 - \omega^2\phi^2 + b' - \frac{\xi}{\phi^2}\right\}\,.
\end{equation}
The remaining field $\phi(t)$ is a time-dependent scale factor determining the area of the spatial slice at time $t$: $V_2(t) = 4\pi\phi^2(t)$.
The constant $b'$ term is clearly irrelevant for the problem of minimization of the action, and this can be traced back to the topological nature of the $R$ term in \eqref{eq:hl-action} for the projectable case.

The good sign of the action (a positive definite kinetic term) corresponds to $\kappa^2>0$, i.e.\ $\lambda <1/2$.
However, in order to satisfy the periodic boundary conditions and to get real oscillating solutions, it turns out that one needs to take $\omega^2 > 0$ and $\xi > 0$ (that is, an  $R^2$ term with the bad sign), thus leading to a potential which is unbounded from below. Both sources of unboundedness are cured by (again CDT-inspired) constraints on the configuration space:

\begin{enumerate}
\item The unbounded $\phi^2$ term is tamed by the the fact that we are at fixed total volume, i.e.\ in the canonical ensemble. As the original $\L$ above, $\omega^2$ should be treated as a Lagrange multiplier.  The Euler-Lagrange equations for $\phi$ are unchanged, while variation with respect to the Lagrange multiplier imposes the constraint:
\begin{equation}
  \label{eq:hl-volume-constraint}
  \mathcal{V} \equiv 4\pi N\int_{-\frac\tau2}^{\frac\tau2} \extd t \, \phi^2(t) - V_3 = 0\,,
\end{equation}
that is, it fixes the 3-volume, and thus there is no unboundedness problem from the sign of the $\omega^2$ term.

  \item The potential unboundedness stemming from the $1/\phi^2$ term is instead avoided by imposing a minimal spatial volume constraint at the outset, such as:\footnote{Note that the compatibility of \eqref{eq:hl-volume-constraint} with \eqref{eq:minimal-constraint} requires $\epsilon<\sqrt{ V_3/(4\pi N\tau) }$.}
\begin{equation}
  \label{eq:minimal-constraint}
  \phi(t) >\epsilon\,,\quad \forall t\,.
\end{equation}
This can be thought as a regularization of the theory, and as discussed in  \cite{Benedetti:2014dra}, there are possible scaling limits to safely let $\epsilon\to 0$.

Alternatively, we could think of the $1/\phi^2$ term in \eqref{eq:hl-mini-action} as a truncation of the large-$\phi$ expansion of a potential that is not singular at $\phi=0$, e.g.\ $1/(\eps^2+\phi^2)$.

\end{enumerate}

Note that the constant lapse $\cN$ has been neatly hidden away in  \eqref{eq:hl-mini-action}. This is rather innocuous in a classical setting. But in principle, it is a degree of freedom that should be integrated over in a quantum regime leading to the imposition of a Hamiltonian constraint. Following \cite{Benedetti:2014dra}, we shall not consider this scenario, rather setting $\cN=1$ from here on. As we discussed above, that is appropriate when wishing to compare to CDT with fixed total time.

In order to rewrite, in analogy to the two and four-dimensional cases, the effective action in terms of area of the slices, we can simply change variables to:
\be
V_2 = 4\pi \phi^2\,,
\ee
which comes from having assumed the slices to be 2-spheres of radius $\phi$.
Leaving aside the constant term and the harmonic term, which is part of the volume constraint, one can rewrite \eqref{eq:hl-mini-action} as:
\be \label{eq:action-eff}
 \tilde S_{\textrm{3d-HL-mini}}[\phi] \equiv \frac{1}{2\kappa^2}\int_{-\frac{\tau}{2}}^{\frac{\tau}{2}} \extd t \left\{\dot{\phi}^2  - \frac{\xi}{\phi^2}\right\} =  \frac{1}{2\kappa^2}\int_{-\frac{\tau}{2}}^{\frac{\tau}{2}} \extd t \left\{\frac{1}{16\pi}\frac{\dot{V_2}^2}{V_2}  - \frac{4\pi\xi}{V_2}\right\}\, \, .
\ee

\textbf{BIB model and its phases:} 
The action \eqref{eq:action-eff} is the form of the action that can most naturally be translated into a discrete BIB model of the type \eqref{Z_BIB}.
The obvious discretization amounts to choosing the following reduced transfer matrix:
\begin{equation}
  \label{eq:bib-2plus1}
  g(m_j,m_{j+1}) = \exp\left\{-\frac{2(m_{j+1} - m_j)^2}{m_{j+1} + m_j}b_1 + \frac{2}{m_{j+1} + m_j}b_2\right\}\,,
\end{equation}
where $b_1$ and $b_2$ are parameters that can be related to the continuous ones, and the factors 2 remind us that we have chosen to write in the denominators the arithmetic mean of $m_j$ and $m_{j+1}$ (other choices, even non-symmetric, are possible, but are expected to be irrelevant in the continuum limit \cite{Bogacz:2012sa}).

%%%%% FIGURE %%%%%%

\begin{figure}[htb]
    \centering
    \includegraphics[scale = 0.5]{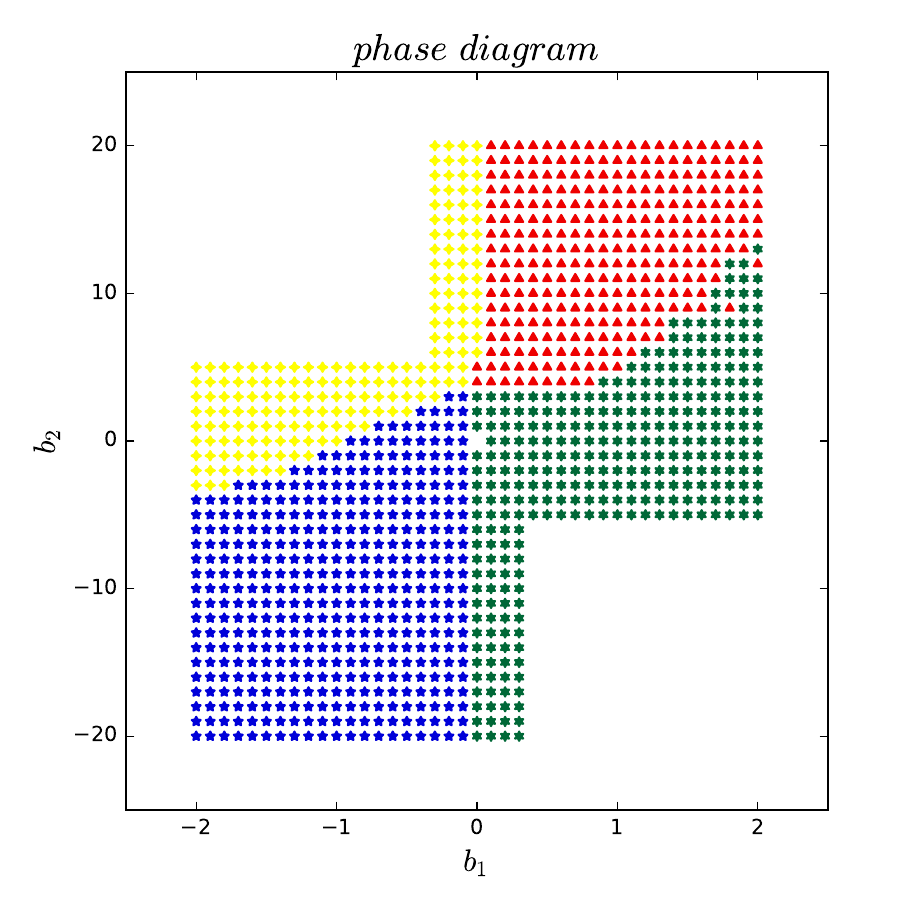}
    \caption{\label{fig:ni-phase} \small{Phase diagram for the BIB model with reduced transfer matrix \eqref{eq:bib-2plus1}, with $T = 80$ and $M = 4000$: droplet (red triangles), localized (yellow squares), antiferromagnetic (blue pentagons), correlated fluid (green hexagons), and uncorrelated fluid (origin).}}
\end{figure}

%%%%% FIGURE %%%%%%

\begin{figure}
    \begin{center}
        \begin{minipage}{0.4\textwidth}
            \centering
            \includegraphics[scale = 0.35]{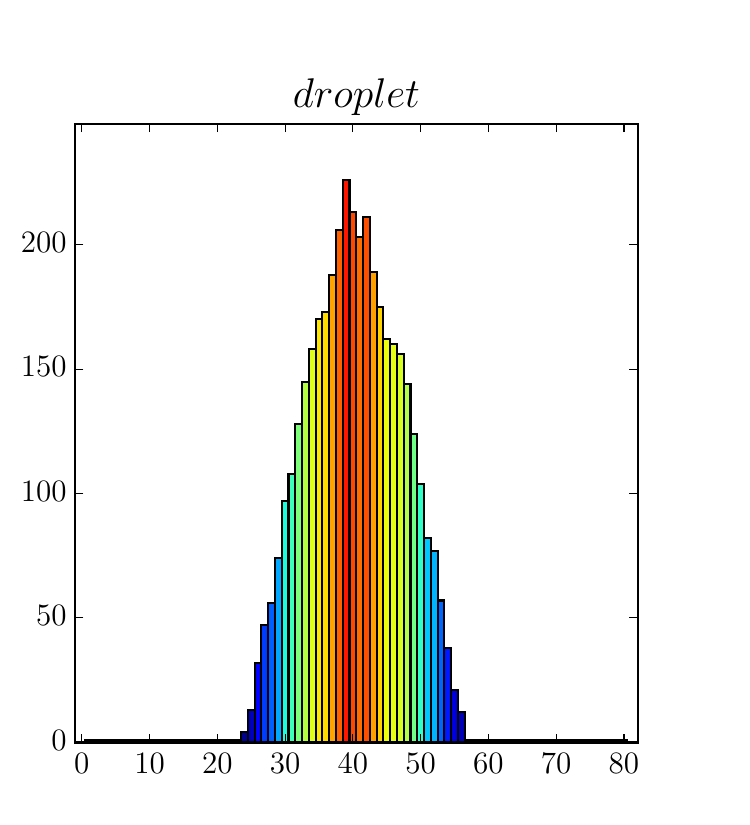}
        \end{minipage}
        \hspace{0.02\textwidth}
        \begin{minipage}{0.4\textwidth}
            \centering
            \includegraphics[scale = 0.35]{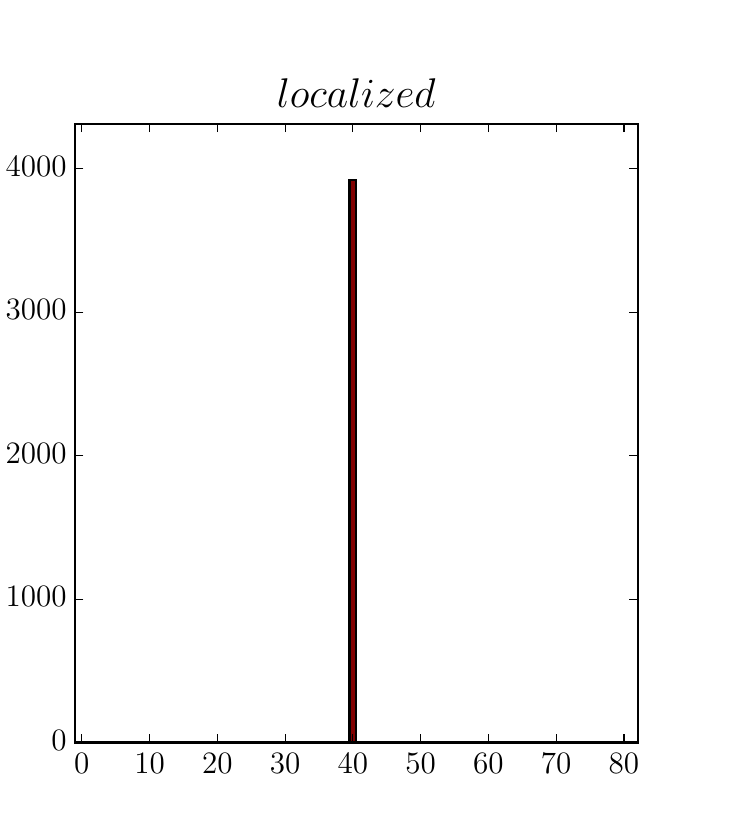}
        \end{minipage}
%        \hspace{0.02\textwidth}
        \begin{minipage}{0.4\textwidth}
            \centering
            \includegraphics[scale = 0.35]{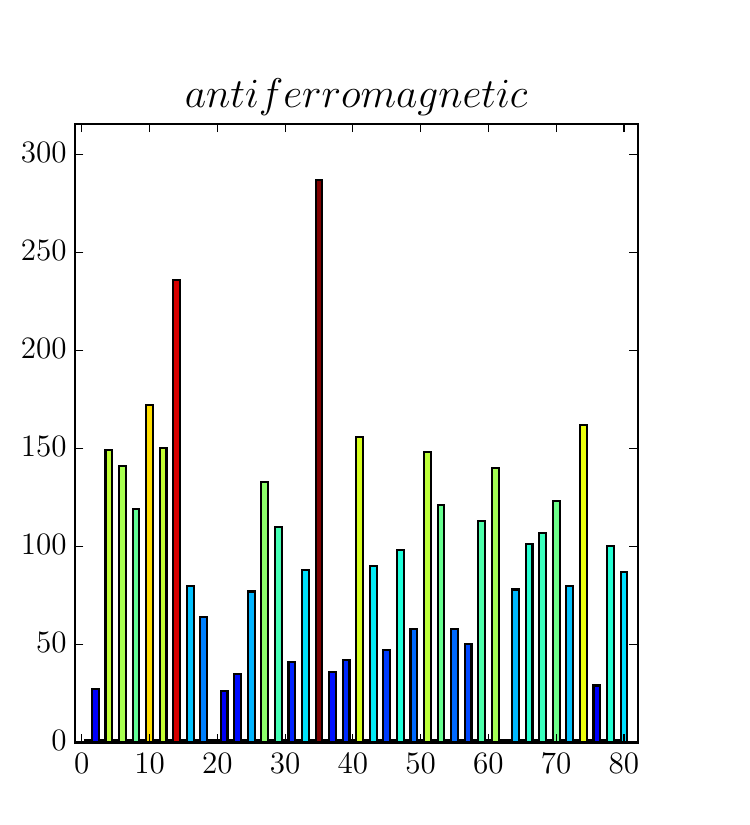}
        \end{minipage}
        \hspace{0.02\textwidth}
        \begin{minipage}{0.4\textwidth}
            \centering
            \includegraphics[scale = 0.35]{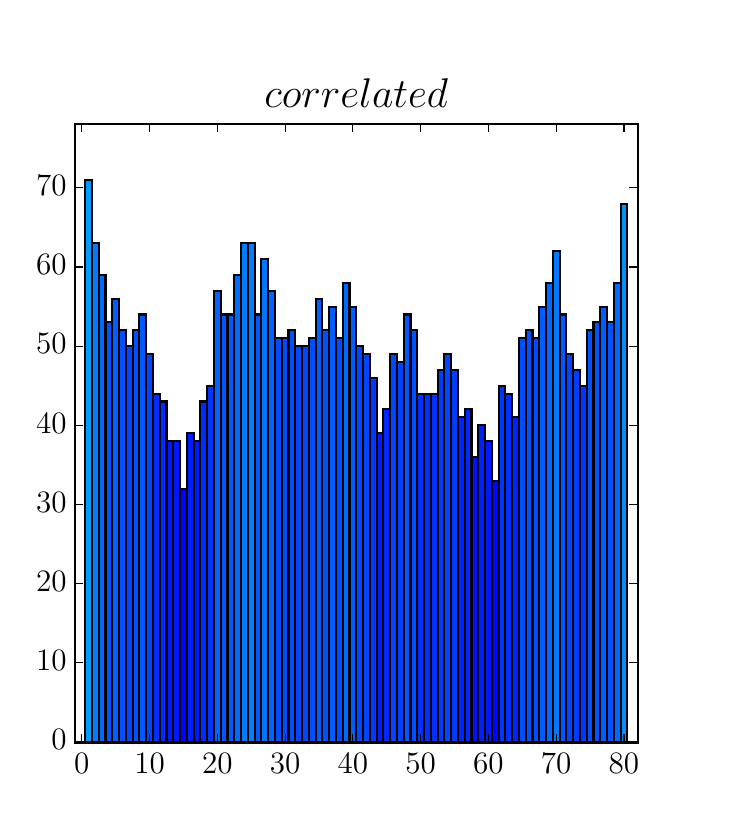}
        \end{minipage}
        \hspace{0.02\textwidth}
        \begin{minipage}{0.4\textwidth}
            \centering
            \includegraphics[scale = 0.35]{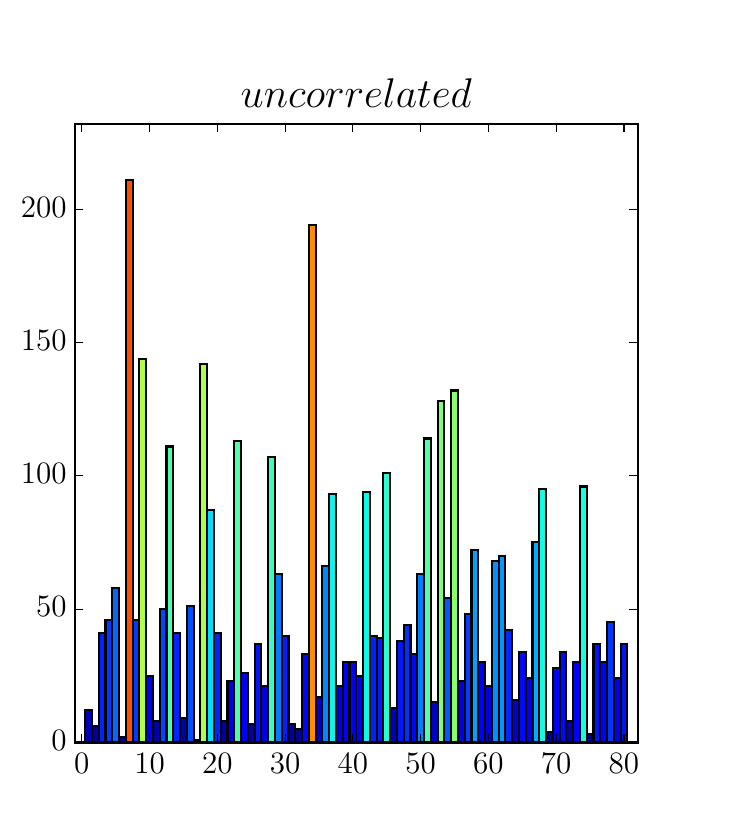}
        \end{minipage}
    \caption{\label{fig:ni-typical} \small{Typical configurations for the various phases (left-to-right): droplet, localized, antiferromagnetic, correlated fluid, uncorrelated fluid.}}
    \end{center}
\end{figure}

We performed Monte Carlo simulations of this BIB model in \cite{Benedetti:2016rwo}, and 
we found evidence for five phases, whose location in the $(b_1,b_2)$ plane is illustrated in figure \ref{fig:ni-phase}. Typical configurations for each phase are presented in figure \ref{fig:ni-typical}, and they are characterized as follows (see \cite{Benedetti:2016rwo} for more details):
\begin{itemize}
    \item {\bf Droplet phase:} This phase occurs for $b_1 > 0$, $b_2 > b_{2,crit}(b_1)$.  For a typical configuration,  the majority of volume condenses onto a contiguous subset of time slices of width greater than one. The remainder of the volume forms a stalk with each constituent time slice possessing minimal volume.
    \item {\bf Localized phase:} This phase exists for $b_1 < 0$, $b_2 > b_{2,crit}(b_1)$. For a typical configuration, all volume resides on one time slice, while all other times possess minimal volume.
    \item {\bf Antiferromagnetic fluid phase:} This phase exists for $b_1 < 0$, $b_2 < b_{2,crit}(b_1)$. A typical configuration consists of alternating peaks and troughs, where the troughs possess minimal volume.
    \item {\bf Correlated fluid phase:} This phase exists for $b_1 > 0$, $b_2 < b_{2,crit}(b_1)$. A typical configuration consists of the volume approximately evenly allotted to the time slices, with relatively small fluctuations about the mean volume.
    \item {\bf Uncorrelated fluid phase:} This phase exists near the origin, i.e. for $b_1\sim b_2 \sim 0$, and probably only at the origin in the thermodynamic limit.
    It is a phase dominated by the entropy of configurations rather than by the action, which indeed vanishes at the origin, where the model is easily solved \cite{Bogacz:2012sa}.
\end{itemize}

%%%%% FIGURE %%%%%%

\begin{figure}
    \begin{center}
        \begin{minipage}{0.45\textwidth}
            \centering
            \includegraphics[scale = 0.35]{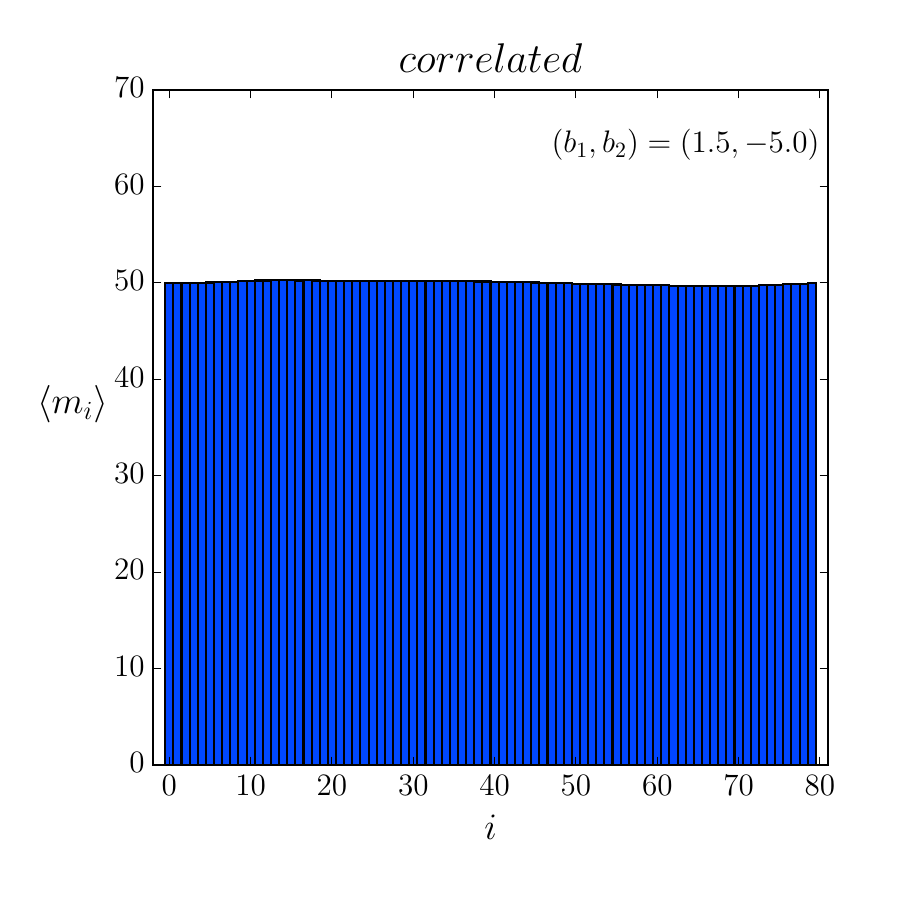}
        \end{minipage}
        \hspace{0.02\textwidth}
        \begin{minipage}{0.45\textwidth}
            \centering
            \includegraphics[scale = 0.35]{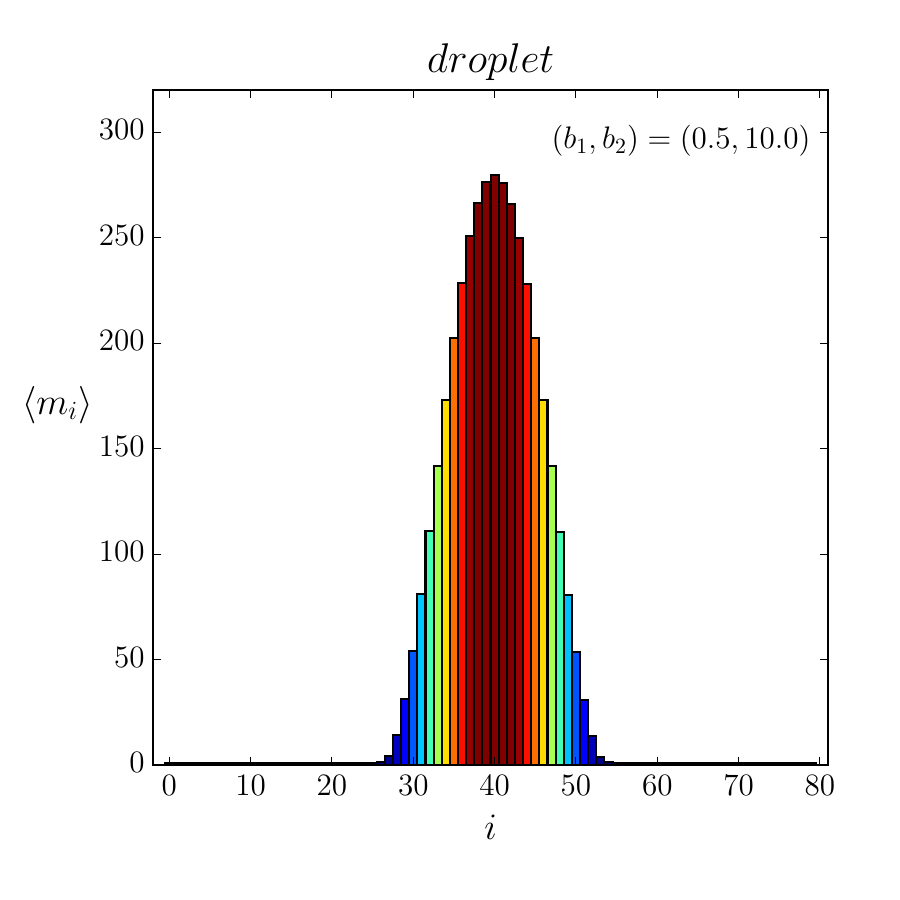}
        \end{minipage}
        \caption{\label{fig:corr-drop_avg} \small{Mean value $\langle m_i\rangle$ as a function of $i$, for non-centred samples in the correlated phase and centred samples in the droplet phase.}}
    \end{center}
\end{figure}

%%%%% FIGURE %%%%%%

Two of the phases above are akin to the two phases appearing in three-dimensional CDT, namely the droplet and the correlated phase. Clearly the former (see its averaged profile in Fig.~\ref{fig:corr-drop_avg}) should correspond to the droplet configuration of Fig.~\ref{Fig:N_2}, and we will get back to their comparison below.
Notice also that although the typical configuration in the correlated phase shows relatively large fluctuation at our value of $M$ the ensemble average in figure \ref{fig:corr-drop_avg} shows that they neatly average to a constant configuration.
The antiferromagnetic and localized phases appear at negative kinetic term, and for this reason we will not discuss them further.

%%%%%%%%%%%%%%%%%%%%%%%%%
\subsection{Analysis of the model}
\label{sec:analysis}
%%%%%%%%%%%%%%%%%%%%%%%%%

\begin{figure}[h]
    \begin{center}
        \includegraphics[width=\textwidth]{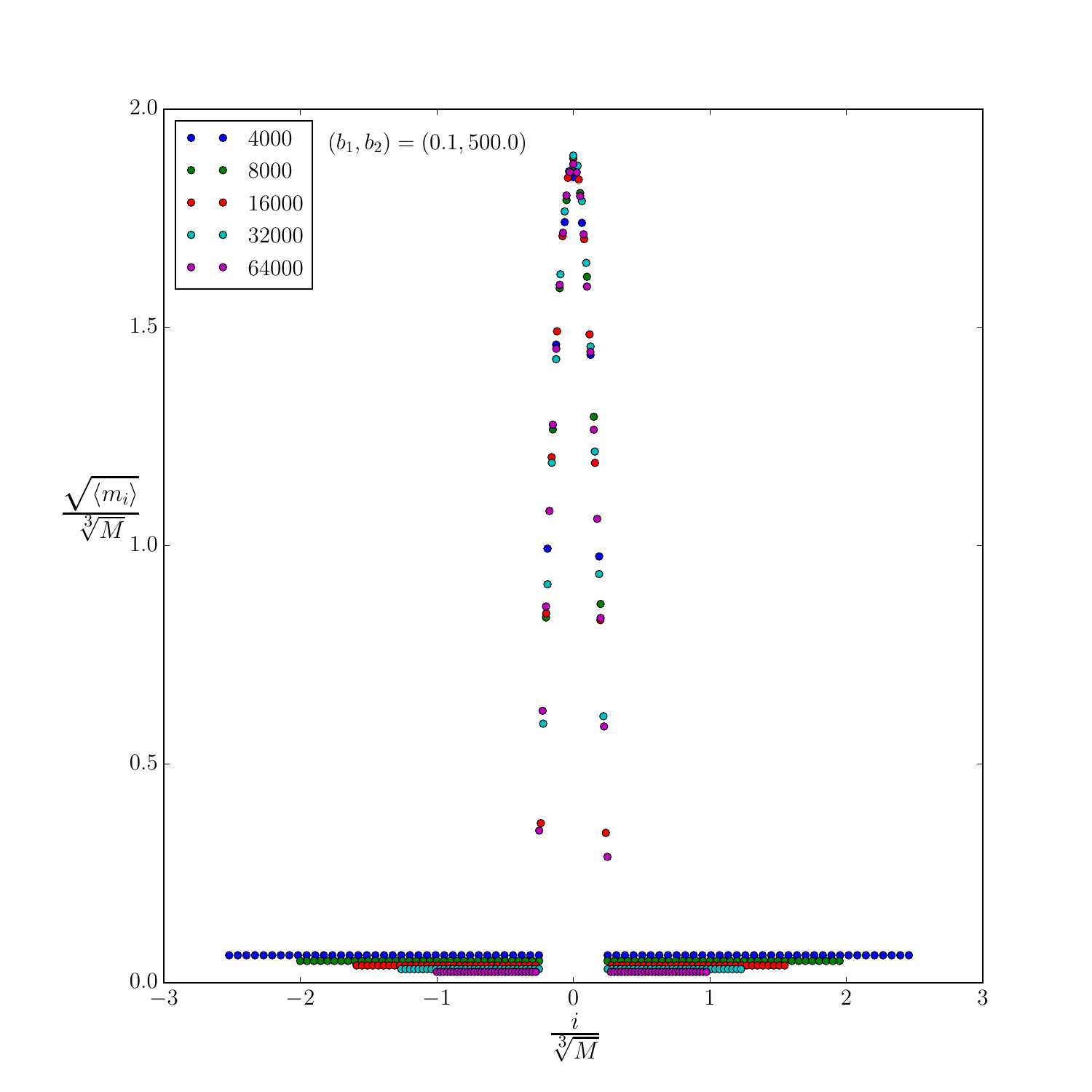}
        \caption{\label{fig:scaling} \small{Scaling of averaged droplet with respect to volume $M$ (between 4k and 64k) at fixed $T = 80$. We notice that rescaling both $\sqrt{\langle m_i\rangle}$ and $i$ with $\sqrt[3]{M}$, we obtain a droplet whose shape is essentially independent of $M$.}}
    \end{center}
\end{figure}

We can reverse the derivation of the BIB model and go back to the continuum, in order to look for the configurations that minimize the Landau free energy.
First off, we notice that a three-dimensional interpretation of the BIB model is justified by the scaling observed in the droplet phase, see Fig.~\ref{fig:scaling}, which is of the same type as the scaling in the three-dimensional CDT model, see Fig.~\ref{Fig:scaling}.

Let us introduce a lattice spacing $a$, and define a continuous time variable  $t= a j$, with period $\tau=a T$. Furthermore, we define a continuous volume variable  $V_3=v_3 a^3 M$, as well as interpret the number of balls in a given box $j$ as the two-dimensional volume of a slice at time $t$: $V_2= v_2 a^2 m$.
Here, $v_2$ and $v_3$ are two arbitrary constants, that could be related to the details of the three-dimensional building blocks of CDT.

In the limit $a\to 0$, with $\tau$ and $V_3$ fixed, the exponent in \eqref{eq:bib-2plus1} becomes the action
\be \label{eq:bib-contS}
S[V_2] = \frac{1}{a} \int_0^\tau dt \left\{ \frac{b_1}{v_2} \frac{\dot{V_2}^2}{V_2} - a^2 b_2 v_2 \frac{1}{V_2} \right\} \, ,
\ee
while the constraint on the total number of balls becomes the following volume constraint:
\be
\int_0^\tau dt V_2 = \frac{v_2}{v_3} V_3 \, .
\ee
It is then natural to take $v_2=v_3\equiv v$.

The presence of the $1/a$ factor in front of the action \eqref{eq:bib-contS} suggests that in the continuum limit the minimization of the action should provide a trustable approach to understanding the phase diagram of the BIB model. Note that it would be wrong to discard the potential term because of the $a^2$ multiplying it, because $V_2$ itself can be of order $a^2$.

A comparison of the actions \eqref{eq:action-eff} and \eqref{eq:bib-contS}  leads to the identifications:
\be
\frac{1}{32\pi\kappa^2} = \frac{b_1}{a v} \,, \;\;\;   \frac{2\pi\xi}{\kappa^2}= a b_2 v \, ,
\ee
or:
\be
\frac{b_2}{b_1} = \frac{64 \pi^2 \xi}{v^2 a^2} \, .
\ee
One can eliminate any reference to the cutoff scale $a$ by looking only at dimensionless ratios of the parameters of the model. For example, using $v a^2 m_{min} = 4\pi \epsilon^2$, one can write:
\be
\frac{v\, b_2}{b_1 m_{min}} = \frac{16 \pi \xi}{\epsilon^2} \, .
\ee
In this way we can translate predictions from the continuum to the discrete.

For the continuum analysis, we will stick to the action in the $\phi$ parametrization, written as in \eqref{eq:hl-mini-action} for deriving the equations of motion, and as in  \eqref{eq:action-eff} for evaluating it on-shell (the $\omega^2$ term being part of the constraint it does not contribute to the on-shell action).
The equations of motion derived from \eqref{eq:hl-mini-action} are of Ermakov-Pinney type, i.e.:
  \begin{equation}
    \label{eq:hl-pe-equations}
    \ddot\phi + \omega^2\phi - \frac{\xi}{\phi^3} = 0\,,
  \end{equation}
and their solution is:
\begin{equation}
  \label{eq:hl-pe-solutions}
  \phi_0(t) = \frac{1}{\omega A} \sqrt{(\omega^2A^4 - \xi)\cos^2(\omega t + \psi) + \xi}\,,
\end{equation}
where $A$ and $\psi$ are integration constants. Shifting the maximum of the curve to $t = 0$ fixes $\psi = 0$, and the maximal value $\phi_0(0) = A$ is then fixed by initial conditions.  Meanwhile, the constraint $\mathcal{V}$ suffices to determine $\omega$ in terms of $\{V_3, A, \xi\}$. However, notice that these solutions oscillate with period $\frac{\pi}{\omega}$, and therefore in order to satisfy periodic boundary conditions at $t=\pm \tau/2$, $\omega$ must also satisfy  $\frac{\pi}{\omega} = \frac{\tau}{n}$ for some positive integer $n$. As a consequence, the space of solutions forms a discrete set.

These solutions never reach zero for $\xi > 0$. This has ramifications for the space-time topology.  In particular, it indicates the non-occurence of potential conical singularities, rather there is a minimal throat, at which the space-time bounces.

Besides the oscillating solutions, there is also a constant solution:
\begin{equation}
  \label{eq:hl-const-solution}
  \bar\phi_0(t) 
  =\sqrt{ \frac{V_3}{4\pi \tau} }\,,
\end{equation}
which is a special case of equation \eqref{eq:hl-pe-solutions}  with $A = \xi^{1/4}/\sqrt{\omega}$, and $\omega=4\pi\tau\sqrt{\xi}/V_3$ fixed by the volume constraint. This solution has special significance, since in comparison to the other local extrema, it has least action:\footnote{The on-shell action evaluates to:
  \begin{equation}
    \label{eq:hl-on-shell-explicit}
    \begin{array}{rcl}
      \tilde S_{\textrm{3d-HL-mini}}[\phi_0] &=&  \dfrac{n\pi}{8\kappa^2}\left(\dfrac{nV_3}{N\tau^2} - 8\sqrt{\xi}\right)\,,\\[0.5cm]
 \tilde S_{\textrm{3d-HL-mini}}[\bar\phi_0]     &=& -\dfrac{2\pi N\tau^2\xi}{\kappa^2V_3}\;.
\end{array}
\end{equation}
}
\begin{equation}
  \label{eq:hl-on-shell}
 \tilde S_{\textrm{3d-HL-mini}}[\bar\phi_0] \leq  \tilde S_{\textrm{3d-HL-mini}}[\phi_0] \, .
\end{equation}

However, a standard second derivative test for constrained local extrema reveals that even the constant solution might not be a stable minimum for $\xi>0$ sufficiently large. Indeed the Hessian of \eqref{eq:hl-mini-action} reads
\be
\cH(t,t') = \left(-\p_t^2 -\om^2 - \f{3\xi}{\phi(t)^4}\right)\d(t-t') \,,
\ee
and on the constant configuration it is diagonalized by the Fourier basis; as the tangent space to the constraint surface \eqref{eq:hl-volume-constraint} at the constant solution is spanned by all the non-constant Fourier modes, we simply find that the condition for such a solution to be a minimum is
\be
\left(\f{2\pi n}{\t}\right)^2  -\om^2 - \f{3\xi}{\bar\phi_0^4} = \left(\f{2\pi n}{\t}\right)^2  - \f{4\xi (4\pi\t)^2}{V_3{}^2}  >0 \,,
\ee
which is violated (at least by the $n=1$ mode) as soon as $\xi>V_3{}^2/(16\t^4)$.

Keeping in mind that droplet-stalk configurations are of interest to us, we notice that the solutions \eqref{eq:hl-pe-solutions} and \eqref{eq:hl-const-solution} above capture fairly well the bulk of the droplet and the stalk, respectively. Therefore, with a touch of naivet\'e, let us graft the two together, and construct a configuration of the form:
\begin{equation}
  \label{eq:hl-graft}
  \tilde\phi(t) = \left\{
    \begin{array}{lcl}
      \dfrac{1}{\omega A}\sqrt{(\omega^2A^4 - \xi)\cos^2(\omega t) + \xi} & \textrm{for} & t \in [-\frac{\pi}{2\omega},+\frac{\pi}{2\omega}]\,, \\[0.3cm]
      \dfrac{\sqrt{\xi}}{\omega A} & \textrm{for} & t\in[-\frac{\tau}{2},-\frac{\pi}{2\omega}) \cup (+\frac{\pi}{2\omega},+\frac{\tau}{2}]\,, \\
    \end{array}
      \right.
\end{equation}
where one period of a configuration of the form \eqref{eq:hl-pe-solutions} is confined to a subinterval of length $\pi/\omega < \tau$, while a constant configuration is attached in the remainder.

Such configurations are unusual for at least two reasons. First, they do not in general correspond to solutions of the Euler-Lagrange equations, except in the degenerate cases $A^2=(V_3+\sqrt{V_3^2-16\tau^4\xi})/(4\pi \tau)$, for which $\omega=\pi/\tau$, i.e. there is no stalk, and $A^2 = \xi^{1/2}/\omega=V_3/(4\pi \tau)$, for which $\tilde\phi = \bar\phi_0$, i.e. there is no droplet. Second, they are only $C^1([-\tau/2,\tau/2])$, the second derivative being discontinuous at $t=\pm \frac{\pi}{2\omega}$.
Nevertheless, we will now argue that they are relevant for the analysis of the model.

\textbf{Local extrema vs absolute minima:}
The partition function of the system, in the continuum, reads:
\begin{equation}
  \label{eq:hl-mini-pf}
  Z_{\textrm{3d-HL-mini}} = \int_{\phi(t)>\epsilon}\mathcal{D}\phi(t)\,
                                \delta(\mathcal{V})\,
                                \exp\left\{
                                -\frac{1}{2\kappa^2}
                                \int_{-\frac\tau2}^{\frac\tau2}\extd t
                                \left[\dot\phi^2 - \frac{\xi}{\phi^2}\right]
                                \right\}\,.
\end{equation}

In the limit $\kappa\to 0$ we expect the partition function (and the observables) to be dominated by those configurations that minimize the action.
Under the assumption that the space of field configurations is such that the action is at least differentiable (in the functional sense), there are two possible ways minima can arise: local extrema (i.e. solutions of the equations of motion) or minima lying on the boundary of configuration space.  In the case that local extrema provide the dominant configurations for the path integral, this would imply dominance of the constant solution, because as we have seen this has the least action among all the solutions. On the other hand, if a configuration lying on the boundary of configuration space provides an absolute minimum, then we can have a non-constant profile.  Such a possibility arises quite naturally in situations where an action unbounded from below is tamed by constraints.
In our case, as we have already pointed out, the unboundedness of the action \eqref{eq:action-eff} at $\xi>0$ is tamed by the constraint \eqref{eq:minimal-constraint}, which introduces a boundary in configuration space. Therefore, absolute minima that are not local extrema can be expected.
Indeed, it was argued in \cite{Benedetti:2014dra} that this is precisely what gives rise to the droplet phase, with a profile of the form \eqref{eq:hl-graft} dominating the partition function.

The rationale behind  \eqref{eq:hl-graft} is based on a balance between kinetic and potential terms in the action (created by their relative sign difference). If one looks at the kinetic term alone (at positive $b_1$, i.e. $\kappa^2$), then it is obvious that the minimizing configuration is a constant one, $\dot\phi(t)=0$. The volume constraint then fixes the value of $\phi(t)$ to \eqref{eq:hl-const-solution}.
On the other hand, if one looks only at the potential term, then minimization would seem to push to a configuration with $\phi(t)=0$, leading to a singular value of the action which would clearly dominate over any other configuration (as long as $\dot\phi(t)$ stays finite).  This is of course the instability caused by the unboundedness of the action, and as already repeatedly stressed it is cured by the constraint \eqref{eq:minimal-constraint}, as a consequence of which the potential actually favors the configuration $\phi(t)=\epsilon$. In fact, the volume constraint forbids a solution with $\phi(t)=\epsilon$ everywhere, but clearly if it were just for the potential term a delta-function configuration $V_2(t)= (V_3-4\pi \epsilon^2\tau)\,  \delta(t-t_0) + 4\pi \epsilon^2$ would do the job,\footnote{Our choice of working with the variable $\phi$ rather than $V_2$ would not be the best in this case due to the need of taking the square root of the delta function.} i.e.\ a localized configuration as seen in the top-right of Fig.~\ref{fig:ni-typical}, which indeed is found in particular at $b_2>0$ and $b_1=0$.

Therefore, for $\epsilon$ small enough, one sees a competition between the kinetic term, which favors the constant configuration, and the potential, which favors the delta-function configuration. What  \eqref{eq:hl-graft} envisages is a situation where the latter essentially wins,  with the constant part being fixed at the minimal slice volume for the optimal choice of $A$, but with the kinetic term smoothing out the delta function, to a shape determined by local minimization of the action.\footnote{Notice that for negative $b_1$ the kinetic term looses its smoothing effect and we might expect the delta function configuration to dominate. This is indeed what we find in the simulations.}

\textbf{Phase transition:}
The discussion above was intended to provide a simple intuition of why a configuration such as \eqref{eq:hl-graft} might dominate the path integral, but of course one has to explicitly check whether that is the case. The presence of several parameters and the fact that \eqref{eq:hl-volume-constraint} leads to a cubic equation for $\omega$ complicate things, and in \cite{Benedetti:2014dra} only a perturbative analysis for small $\epsilon$ and $\xi$ was presented.

Indeed plugging \eqref{eq:hl-graft} into \eqref{eq:hl-volume-constraint} we find:
\be \label{V3constr}
V_3 = 4\pi  \left(  \f{ \pi (\xi+A^4\omega^2) }{2 A^2 \omega^3} +\f{\xi}{A^2\omega^2} \left(\tau-\f{\pi}{\omega}\right)\right)\, ,
\ee
which, multiplied by $\omega^3$ (non-zero otherwise we would have the constant solution again), gives us a cubic equation for $\omega$. It is convenient to rewrite the latter in terms of $\epsilon$ rather than $A$, taking $A=\sqrt{\xi}/(\epsilon\omega)$, because we expect the constant part to reach its minimal allowed value (from the considerations above, and from the perturbative analysis of \cite{Benedetti:2014dra}). We arrive at:
\be \label{eq:cubic-omega}
(4\pi \tau \epsilon^4-V_3\epsilon^2) \omega^3 -2\pi^2\epsilon^4 \omega^2+2\pi^2 \xi = 0\, .
\ee
Although solvable, the general solution to such an equation is not very enlightening due to the presence of several parameters.
We can gain some insight with some further assumptions on the nature of the roots. The discriminant of the cubic equation is:
\be
-108 \pi^4 \xi^2 \epsilon^4 \left(V_3-4 \pi  \tau \epsilon^2\right)^2+64 \pi^8 \xi  \epsilon^{12} \, ,
\ee
and it has the following two roots when viewed as a function of $\xi$:
\be
\xi_1=0\, , \;\;\; \text{and} \;\;\; \xi_2 = \frac{16 \pi^4 \epsilon^8}{27 \left(V_3-4 \pi  \tau \epsilon^2\right)^2} >0 \, .
\ee
For small $\epsilon$ and large $V_3$, the positive root takes a very small value, and therefore we concentrate on the case $\xi>\xi_2$. In such case, the discriminant is negative and therefore the cubic equation has only one real root. Then we can use the representation of the real root in terms of hyperbolic functions, writing:
\be \label{eq:omega-sol}
\omega_0 = 2 \sqrt{\frac{-p}{3}} \cosh \left(\frac{1}{3} {\rm arcosh} \left(\frac{3 q}{2 p}
   \sqrt{-\frac{3}{p}}
   \right)\right)-\frac{a_2}{3 a_3} \, ,
\ee
where:
\be
p = \frac{3 a_3 a_1-a_2^2}{3 a_3^2} = -\frac{4 \pi^4 \epsilon^4}{3 \left(V_3-4 \pi  \tau \epsilon^2\right)^2} \, ,
\ee
\be
q = \frac{27 a_3^2 a_0-9 a_3 a_2 a_1+2 a_2^3}{27 a_3^3} = \frac{16 \pi ^6 \epsilon ^{8}-54 \pi ^2 \xi
 \left(V_3-4 \pi  \tau \epsilon^2\right)^2}{27\epsilon^2 \left(V_3-4 \pi  \tau \epsilon^2\right)^3}\, ,
\ee
and $a_n$ is the coefficient of $\omega^n$ in the cubic equation \eqref{eq:cubic-omega}. Furthermore, in \eqref{eq:omega-sol} we have assumed $p<0$, $q<0$ and $4 p^3 +27 q^2>0$, which are all valid for large enough $V_3$.

Lastly, we take \eqref{eq:hl-graft} with  $A=\sqrt{\xi}/(\epsilon\omega)$ and replace $\omega$ by the solution \eqref{eq:omega-sol}, to obtain a profile which is a function of $\xi$, $V_3$, $\tau$ and $\epsilon$.
Denoting such configuration as $\hat \phi_0(t)$, we are interested in studying:
\be
\mathcal S \equiv \tilde S_{\textrm{3d-HL-mini}}[\bar\phi_0] - \tilde S_{\textrm{3d-HL-mini}}[\hat\phi_0]
\ee
as a function of $\xi$, at fixed $\kappa^2$, $V_3$, $\tau$ and $\epsilon$, in order to check if and when the droplet configuration \eqref{eq:hl-graft} dominates  (i.e. $\mathcal S>0$).

More conveniently, we can re-express $\mathcal S$ in terms of the discrete BIB dimensionless parameters, eliminating $\xi$, $V_3$ and $\tau$ by means of the relations:
\be
\frac{16 \pi \xi}{\epsilon^2} = \frac{v b_2}{b_1 m_{min}}\, ,
\ee
\be
\frac{V_3}{(4\pi \epsilon^2)^{3/2}} = \frac{M}{m_{min}^{3/2} v^{1/2}} \, ,
\ee
\be
\frac{\tau}{(4\pi \epsilon^2)^{1/2}} = \frac{T}{ (m_{min} v)^{1/2}} \, .
\ee
One then finds that both $\epsilon$ and $\sqrt{v}$ factor out, so that $\mathcal{S}'\equiv  \kappa^2\mathcal{S}/(\epsilon\sqrt{v})$ depends only on the remaining discrete parameters. In particular, for $m_{min}=1$,  $\mathcal{S}'$ depends only on $M$, $T$ and the ratio $b_2/b_1$.

We thus come to our main conclusion: for fixed $M$ and $T$, the reasoning based on the minimization of the action predicts that there is a phase transition between a droplet ($\mathcal S>0$) and a correlated fluid phase ($\mathcal S<0$), with the boundary between the two phases being given by a straight line in the $(b_1,b_2)$ plane.
In fact, for fixed $M$ and $T$, $\mathcal{S}'$ is only a function of the ratio $\alpha\equiv b_2/b_1$, and the point at which  $\mathcal{S}'(\alpha_c)=0$ corresponds to the phase transition.
Unfortunately, such an equation cannot be solved in a closed form, and we are limited to a numerical evaluation of $\alpha_c$.
For example, at $M=4000$ and $T=80$ , we find $\alpha_c=1.5$, while  at $M=16000$ and $T=80$ , we find $\alpha_c=7.1$.
We can also check numerically that at large volume $\alpha_c \propto M/T^3$. However, the quantitative predictions about the location of the phase transition should not be taken too seriously, because near the phase transition we expect the fluctuations to become important and affect the transition point.

The above results show that the droplet configuration \eqref{eq:hl-graft} wins over the constant one in a certain range of parameters. What we have not shown is that there are no other solutions that win over both in that same range of parameters. In fact it is plausible that the actual dominant configuration is a smooth version of  \eqref{eq:hl-graft}, i.e. one in which bulk and stalk are joined smoothly. However, we believe that the mechanism described above for our ansatz correctly captures the essential physical features of the problem, and this feeling is corroborated by the numerical simulations that have confirmed the existence of a droplet phase in the BIB model  \eqref{Z_BIB} with reduced transfer matrix \eqref{eq:bib-2plus1}.

%%%%%%%%%%%%%%%%%%%%%%%%%%
\section{Conclusions}
\label{sec:conclusions}
%%%%%%%%%%%%%%%%%%%%%%%%%%%

In this review we have argued that the spatial volume profile in CDT can be interpreted as a sort of coarse grained order parameter, and that its associated Landau free energy can help us to understand the CDT phased diagram and the dynamics leading to it. At the same time it can also teach us something about the continuum limit, thanks to the partially local nature of such observable.
Armed with such observable, and with the Landau theory point of view in mind, we have made the case that the continuum limit of CDT is most likely described by an FPD-invariant effective field theory, as in HL-type theories.

\begin{figure}
        \centering
        \includegraphics[width=.8\textwidth]{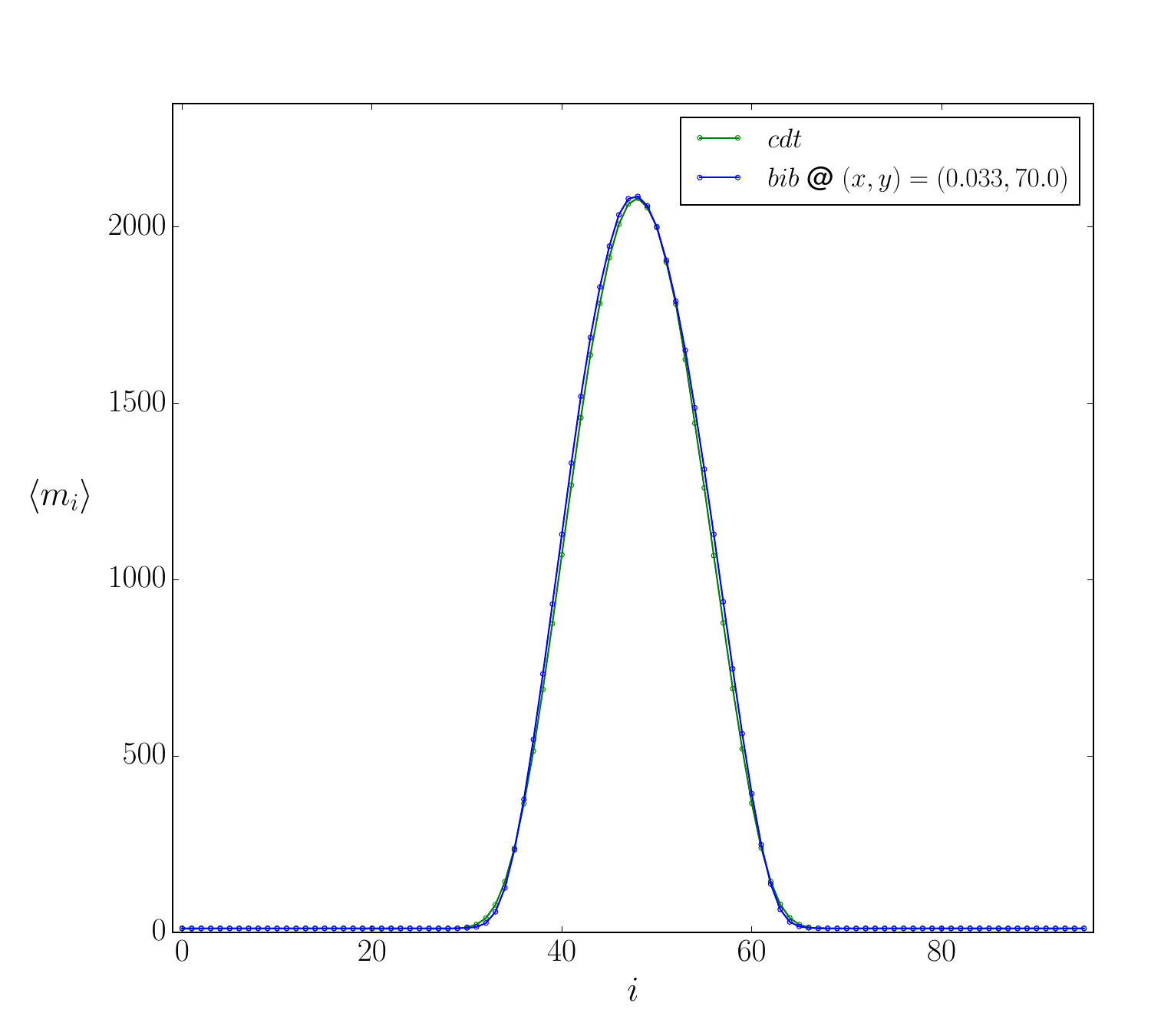}
        \caption{\label{fig:bib-cdt}
        \small{Comparison of the mean volume profile from the BIB model and three-dimensional CDT.}}
\end{figure}

In particular, we have first reviewed the fact that in two-dimensional CDT the Landau free energy for the spatial length is obtained exactly and in the continuum limit it coincides with the two-dimensional HL gravity model.
We have then argued that in order to explain the spacetime condensation phenomenon of three-dimensional CDT, reviewed in Sec.~\ref{Sec:data}, a Landau free energy obtained as the discretization of a minisuperspace reduction of the Einstein-Hilbert action is not sufficient, as it would instead predict the dominance of a constant profile.
By studying instead a Landau free energy inspired by HL gravity, proposed in \cite{Benedetti:2014dra}, we have shown that a continuum semiclassical analysis leads to a condensation, captured by the dominant profile \eqref{eq:hl-graft}, compatible with the one observed in CDT.
Such a semiclassical analysis is corroborated by the Monte Carlo simulations \cite{Benedetti:2016rwo} of the effective BIB model, whose main results are summarized by the figures \ref{fig:ni-phase} and \ref{fig:bib-cdt}. The former shows the rich phase diagram of the BIB model proposed in \cite{Benedetti:2014dra}. In particular, note the presence of a droplet phase, which however would be absent were one to set $b_2=0$, as in the BIB model related to minisuperspace reduction of general relativity. Figure \ref{fig:bib-cdt} shows instead the excellent correspondence between the droplet phase of the BIB model and the CDT volume profile. Thus, we have compelling evidence that the effective dynamics of the CDT spatial slices is well captured by the BIB model \eqref{Z_BIB}-\eqref{eq:bib-2plus1}, which in turn can be seen as a discretized path integral for a minusuperspace reduction of the Ho\v{r}ava-Lifshitz model \eqref{eq:hl-action}.
%%%%%%%%%%%%%%%%%%%%%%%%
%%%%%%%%%%%%%%%%%%%%%%%%

\

As we mentioned along the course of this review, there are a number of other reasons in support of a link between CDT and HL gravity (or more generally,  FPD-invariant effective field theories).
It is worth to collect them here in order to summarize the status of this relation:

\begin{itemize}

\item {\bf Presence of a foliation:} This is on one hand a trivial observation, as it is a defining property of both CDT and HL gravity, and on the other a subtle one, as understanding the fate of the CDT foliation in the conitnuum limit is nontrivial. In particular, it is nontrivial because we have not yet identified a proper tuning towards the second-order phase transition that might lead to a continuum limit with propagating degrees of freedom (see \cite{Ambjorn:2014gsa,Ambjorn:2020rcn} for a first attempt).
As a counterargument to the importance of the foliation, a generalized CDT model with a local causality constraint but no built-in foliation was introduced in \cite{Jordan:2013awa,Jordan:2013iaa} and shown to also lead to a semiclassical droplet phase with similar properties to that of standard CDT. On the other hand, even if such generalized CDT model would turn out to be indeed in the same universality class of usual CDT, this would not tell us what is their continuum interpretation: it might well be that the FPD-invariant universality class is robust under small departures from the foliated structure at the discrete level. 
Moreover, as argued in the introduction, the very fact that observables that are local in time are available and are being studied in CDT is by itself a signal of a likely departure from a full-diffeomorphism invariant theory. 

\item {\bf Analogies in the phase diagram:} This was discussed for example in \cite{Ambjorn:2010hu}, and it is based on the observation that a Lifshitz point is typically a tricritical point  at the intersection of feromagnetic, paramagnetic and modulated phases, and that such phases, with their analog solutions in HL gravity, have a resemblance to the original three phases observed in four-dimensional CDT. 
While the comparison can be based on classical solutions and analogy to the Lifshitz scalar, the most concrete realization is in the kind of analysis we reviewed here: as reviewed in Sec.~\ref{Sec:3d_BIB}, the BIB model based on HL gravity leads to the phase diagram of Fig.~\ref{fig:ni-phase}, which at $b_1>0$ matches the phase diagram of CDT.  A similar result holds also in four dimensions \cite{Bogacz:2012sa}: in this case, no higher-order terms in the curvature are used, but nevertheless the fact that the coefficients of the kinetic and potential terms are tuned independently in order to reach the different phases suggests again a relation to HL gravity, as the ratio between the two would instead be fixed in the Einstein-Hilbert action. 
However, the four-dimensional picture was perturbed by the discovery of a new "bifurcation" phase in CDT \cite{Ambjorn:2014mra,Ambjorn:2016mnn}, which depends on peculiar properties of the three-dimensional slices.
Such phase has no analogue in the BIB model, which is not surprising as it is due to a change in the internal structure of the three-dimensional slices, which cannot be captured by the spatial volume alone. In other words, it requires another order parameter in order to be identified. It has also been suggested that the appearance of the bifurcation phase might not be described by a conventional Landau-type phase transition \cite{Ambjorn:2020rcn}.
This is one more reason why we have focused here on three-dimensional CDT, where no analogue of the bifurcation phase is found.

\item {\bf Short-scale spectral dimension:} The fact that in the droplet phase of CDT the spectral dimension matches the classical value at (not too) large scales, and decreases at short scales \cite{Ambjorn:2005db} was shown to be reproduced by HL gravity in \cite{Horava:2009if}. However, several other approaches to quantum gravity claimed a similar result (e.g.\ in asymptotic safety \cite{Lauscher:2005qz}), thus weakening this point of contact. On the other hand, in \cite{Benedetti:2009ge} it was pointed out that in three dimensions asymptotic safety and HL gravity give different predictions, and numerical evidence was produced in CDT favoring the HL scenario over the asymptotic safety one.

\item {\bf Large-scale geometry:} In \cite{Benedetti:2009ge} it was also observed that the behavior of the spectral dimension at large scales, where it decays due to the compactness of spacetime, is better fitted by a stretched sphere geometry, rather than a round sphere. And the former is a solution to HL gravity.

\item {\bf Absence of conformal factor problem:} As we discussed in Sec.~\ref{sec:BIB}, the effective actions controlling the dynamics of the spatial volumes in CDT in $d=3$ and 4, have a kinetic term with the good sign, that is, there is no conformal sickness.
While perhaps this could be interpreted as being due to nonperturbative contributions from the functional measure, as suggested in \cite{Dasgupta:2001ue}, it might also have the much simpler explanation that in HL gravity the conformal factor problem can be avoided by an appropriate choice of the coupling $\l$ in \eqref{eq:hl-action}.
More investigations are probably needed in order to decide between the two scenarios, but one should keep in mind that if nonperturbative contributions could change the sign of the kinetic term for the conformal mode with respect to usual Euclidean quantum gravity \cite{Gibbons:1978ac}, then either they would also spoil the good sign of the graviton term in the latter, or they would be hard to reconcile with general covariance of the theory, as the relative sign between conformal and graviton terms is fixed by full-diffeomorphism invariance. Unfortunately no results are available in CDT about a possible graviton effective action, but the possibility that only the conformal factor's sign is changed with respect to  usual Euclidean quantum gravity is corroborated by the results from three-dimensional CDT with toroidal slices \cite{Budd:2011zm}, according to which both the conformal factor and the moduli have the good sign.

\item {\bf Potential versus higher-derivatives:}
Results from \cite{Ambjorn:2011ph} indicate that in four-dimensional CDT higher-order time derivatives in the effective action are subleading with respect to potential terms, that in a full action would correspond to higher-order spatial derivatives. This is suggestive again of an anisotropy.

\item {\bf Exact relation in two dimensions:} Lastly, two-dimensional CDT is exactly solvable and in the continuum limit it reproduces HL gravity, as reviewed in Sec.~\ref{sec:2dCDT}, and more in detail in \cite{Sato:2022ory}. 
As there is currently no hint of a possible peculiarity of two dimensions when it comes to the CDT foliation,\footnote{There is of course a difference in the HL action leading to a renormalizable theory, but this is irrelevant for an effective field theory argument.} this makes a strong plausibility argument for a relation between CDT and HL gravity also in higher dimensions.

\end{itemize}

Clearly understanding the nonperturbative dynamics and continuum limit of a statistical model of higher-dimensional random geometries such as CDT is a very challenging task, which explains why many of the points above are still inconclusive and at a conjectural stage. 
However, we find that the while perhaps none of the arguments in support of a FPD-invariant effective field theory is conclusive in itself, the list of arguments is sufficiently long and varied, so that a different interpretation in terms of a fully diffeomorphism-invariant effective field theory would look as a lucky conspiracy.

It should be stressed that clearly CDT in its usual definition has less couplings than renormalizable HL gravity models in three or four dimensions.
Therefore, such renormalizable models are probably not directly relevant for comparisons to CDT, unless one extends the latter with appropriate new terms in the action, as done in \cite{Anderson:2011bj}.
However, as we argued in the introduction and in Sec.~\ref{sec:2dCDT}, the fact that one starts in CDT with a discretization of the Einstein-Hilbert action\footnote{And moreover we remark again that in four dimensions the CDT action has one more coupling than the Einstein-Hilbert action.} does not imply that its symmetries will be preserved. The crucial point, on which we focused in this review by indirect means, is to assess in which theory space a renormalization group flow of CDT will lead us, or in other words what is the effective field theory of CDT.
Once that is established, only in a second step we will be able to worry about how to take the continuum limit with propagating degrees of freedom corresponding to our desired target theory.
It is at this step that the fact that CDT has only few tunable parameters might turn out to be a limitation. At present, the two competing scenarios that can serve as justification for a program like CDT are asymptotic safety and HL gravity. While the latter is based on a Gaussian fixed point, and so it is straightforward to count the number of its relevant couplings,\footnote{It is instead not so straightforward to study its renormalization group flow \cite{Benedetti:2013pya,Barvinsky:2017kob,Barvinsky:2019rwn,Barvinsky:2021ubv}.}
 the former is based on a conjectural non-Gaussian fixed point, and we do not even know for sure if the number of relevant couplings is finite (except in special simple approximations \cite{Benedetti:2013jk,Mitchell:2021qjr}). However, all evidence so far suggests that if such a nontrivial fixed point exists, then it has at least three relevant perturbations, which include higher-order curvature terms \cite{Codello:2008vh,Benedetti:2009rx,Falls:2014tra}.
Therefore, continuum scenarios and their results suggest that at some point one might need to introduce more couplings in CDT.

Lastly, we stress again that if indeed the effective field theory of CDT turns out to be a FPD-invariant one, this does not exclude the existence of a subspace of its theory space which is invariant under full diffeomorphisms. In other words, one might hope to embed asymptotically safe Euclidean gravity in the HL theory space.
Reaching such subspace would probably require again a very delicate fine tuning of several breaking terms, yet perhaps in a less dramatic way than a full breaking of diffeomorphism invariance requires, as in the functional renormalization group approach to asymptotic safety.

\begin{acknowledgement}
I would like to thank Joe Henson and James Ryan for the precious collaboration on the work reviewed here. I also thank Jan Ambjorn and Renate Loll for the invitation to contribute a chapter to the CDT section of the Handbook of Quantum Gravity, and in particular Jan Ambjorn for useful feedback on the draft.
\end{acknowledgement}

%
%\section*{Appendix}
%\addcontentsline{toc}{section}{Appendix}
%
%
%When placed at the end of a chapter or contribution (as opposed to at the end of the book), the numbering of tables, figures, and equations in the appendix section continues on from that in the main text. Hence please \textit{do not} use the \verb|appendix| command when writing an appendix at the end of your chapter or contribution. If there is only one the appendix is designated ``Appendix'', or ``Appendix 1'', or ``Appendix 2'', etc. if there is more than one.

\providecommand{\href}[2]{#2}\begingroup\raggedright\endgroup


\begin{thebibliography}{100}

\bibitem{Aizenman:2019yuo}
M.~Aizenman and H.~Duminil-Copin, \emph{{Marginal triviality of the scaling
  limits of critical 4D Ising and $\phi_4^4$ models}},
  \href{https://doi.org/10.4007/annals.2021.194.1.3}{\emph{Annals Math.}
  {\bfseries 194} (2021) 163}
  [\href{https://arxiv.org/abs/1912.07973}{{\ttfamily 1912.07973}}].

\bibitem{Weinberg:1980gg}
S.~Weinberg, \emph{{ULTRAVIOLET DIVERGENCES IN QUANTUM THEORIES OF
  GRAVITATION}},  in \emph{{General Relativity: An {E}instein Centenary
  Survey}}, S.W.~Hawking and W.~Israel, eds., (Cambridge, UK), pp.~790--831,
  Univ. Pr. (1979).

\bibitem{Niedermaier:2006wt}
M.~Niedermaier and M.~Reuter, \emph{{The Asymptotic Safety Scenario in Quantum
  Gravity}}, \href{https://doi.org/10.12942/lrr-2006-5}{\emph{Living Rev. Rel.}
  {\bfseries 9} (2006) 5}.

\bibitem{Bonanno:2020bil}
A.~Bonanno, A.~Eichhorn, H.~Gies, J.M.~Pawlowski, R.~Percacci, M.~Reuter
  et~al., \emph{{Critical reflections on asymptotically safe gravity}},
  \href{https://doi.org/10.3389/fphy.2020.00269}{\emph{Front. in Phys.}
  {\bfseries 8} (2020) 269} [\href{https://arxiv.org/abs/2004.06810}{{\ttfamily
  2004.06810}}].

\bibitem{Ambjorn:1997di}
J.~Ambj\o{}rn, B.~Durhuus and T.~Jonsson, \emph{{Quantum Geometry}: {A
  Statistical Field Theory Approach}}, Cambridge Monographs on Mathematical
  Physics, Cambridge Univ. Press, Cambridge, UK (12, 2005),
  \href{https://doi.org/10.1017/CBO9780511524417}{10.1017/CBO9780511524417}.

\bibitem{Regge:1961px}
T.~Regge, \emph{General relativity without coordinates},
  \href{https://doi.org/10.1007/BF02733251}{\emph{Nuovo Cim.} {\bfseries 19}
  (1961) 558}.

\bibitem{David:1984tx}
F.~David, \emph{{Planar Diagrams, Two-Dimensional Lattice Gravity and Surface
  Models}}, \href{https://doi.org/10.1016/0550-3213(85)90335-9}{\emph{Nucl.
  Phys. B} {\bfseries 257} (1985) 45}.

\bibitem{Ambjorn:1985az}
J.~Ambjorn, B.~Durhuus and J.~Frohlich, \emph{{Diseases of Triangulated Random
  Surface Models, and Possible Cures}},
  \href{https://doi.org/10.1016/0550-3213(85)90356-6}{\emph{Nucl. Phys. B}
  {\bfseries 257} (1985) 433}.

\bibitem{Kazakov:1985ea}
V.A.~Kazakov, A.A.~Migdal and I.K.~Kostov, \emph{{Critical Properties of
  Randomly Triangulated Planar Random Surfaces}},
  \href{https://doi.org/10.1016/0370-2693(85)90669-0}{\emph{Phys. Lett. B}
  {\bfseries 157} (1985) 295}.

\bibitem{DiFrancesco:1993nw}
P.~Di~Francesco, P.H.~Ginsparg and J.~Zinn-Justin, \emph{{{$2-D$} {Gravity} and
  random matrices}},
  \href{https://doi.org/10.1016/0370-1573(94)00084-G}{\emph{Phys. Rept.}
  {\bfseries 254} (1995) 1}
  [\href{https://arxiv.org/abs/hep-th/9306153}{{\ttfamily hep-th/9306153}}].

\bibitem{Budd:2022zry}
T.~Budd, \emph{{Lessons from the Mathematics of Two-Dimensional Euclidean
  Quantum Gravity}},  in \emph{Handbook of Quantum Gravity}, C.~Bambi,
  L.~Modesto and I.~Shapiro, eds., Springer Singapore (2023)
  [\href{https://arxiv.org/abs/2212.03031}{{\ttfamily 2212.03031}}].

\bibitem{deBakker:1996zx}
B.V.~de~Bakker, \emph{{Further evidence that the transition of 4-D dynamical
  triangulation is first order}},
  \href{https://doi.org/10.1016/S0370-2693(96)01277-4}{\emph{Phys. Lett. B}
  {\bfseries 389} (1996) 238}
  [\href{https://arxiv.org/abs/hep-lat/9603024}{{\ttfamily hep-lat/9603024}}].

\bibitem{Bialas:1996wu}
P.~Bialas, Z.~Burda, A.~Krzywicki and B.~Petersson, \emph{{Focusing on the
  fixed point of 4-D simplicial gravity}},
  \href{https://doi.org/10.1016/0550-3213(96)00214-3}{\emph{Nucl. Phys. B}
  {\bfseries 472} (1996) 293}
  [\href{https://arxiv.org/abs/hep-lat/9601024}{{\ttfamily hep-lat/9601024}}].

\bibitem{Ambjorn:1995dj}
J.~Ambjorn and J.~Jurkiewicz, \emph{{Scaling in four-dimensional quantum
  gravity}}, \href{https://doi.org/10.1016/0550-3213(95)00303-A}{\emph{Nucl.
  Phys. B} {\bfseries 451} (1995) 643}
  [\href{https://arxiv.org/abs/hep-th/9503006}{{\ttfamily hep-th/9503006}}].

\bibitem{Bonzom:2011zz}
V.~Bonzom, R.~Gurau, A.~Riello and V.~Rivasseau, \emph{{Critical behavior of
  colored tensor models in the large N limit}},
  \href{https://doi.org/10.1016/j.nuclphysb.2011.07.022}{\emph{Nucl.\ Phys.\ B}
  {\bfseries 853} (2011) 174}
  [\href{https://arxiv.org/abs/1105.3122}{{\ttfamily 1105.3122}}].

\bibitem{Gurau:book}
R.~Gurau, \emph{{Random Tensors}}, Oxford University Press, Oxford (2016).

\bibitem{Laiho:2011ya}
J.~Laiho and D.~Coumbe, \emph{{Evidence for Asymptotic Safety from Lattice
  Quantum Gravity}},
  \href{https://doi.org/10.1103/PhysRevLett.107.161301}{\emph{Phys. Rev. Lett.}
  {\bfseries 107} (2011) 161301}
  [\href{https://arxiv.org/abs/1104.5505}{{\ttfamily 1104.5505}}].

\bibitem{Benedetti:2011nn}
D.~Benedetti and R.~Gurau, \emph{{Phase Transition in Dually Weighted Colored
  Tensor Models}},
  \href{https://doi.org/10.1016/j.nuclphysb.2011.10.015}{\emph{Nucl.\ Phys.\ B}
  {\bfseries 855} (2012) 420}
  [\href{https://arxiv.org/abs/1108.5389}{{\ttfamily 1108.5389}}].

\bibitem{Gurau:2013cbh}
R.~Gurau and J.P.~Ryan, \emph{{Melons are branched polymers}},
  \href{https://doi.org/10.1007/s00023-013-0291-3}{\emph{Annales Henri
  Poincare} {\bfseries 15} (2014) 2085}
  [\href{https://arxiv.org/abs/1302.4386}{{\ttfamily 1302.4386}}].

\bibitem{Bonzom:2016dwy}
V.~Bonzom, \emph{{Large $N$ Limits in Tensor Models: Towards More Universality
  Classes of Colored Triangulations in Dimension $d\geq 2$}},
  \href{https://doi.org/10.3842/SIGMA.2016.073}{\emph{SIGMA} {\bfseries 12}
  (2016) 073} [\href{https://arxiv.org/abs/1603.03570}{{\ttfamily
  1603.03570}}].

\bibitem{Ambjorn:2013eha}
J.~Ambjorn, L.~Glaser, A.~Goerlich and J.~Jurkiewicz, \emph{{Euclidian 4d
  quantum gravity with a non-trivial measure term}},
  \href{https://doi.org/10.1007/JHEP10(2013)100}{\emph{JHEP} {\bfseries 10}
  (2013) 100} [\href{https://arxiv.org/abs/1307.2270}{{\ttfamily 1307.2270}}].

\bibitem{Coumbe:2014nea}
D.~Coumbe and J.~Laiho, \emph{{Exploring Euclidean Dynamical Triangulations
  with a Non-trivial Measure Term}},
  \href{https://doi.org/10.1007/JHEP04(2015)028}{\emph{JHEP} {\bfseries 04}
  (2015) 028} [\href{https://arxiv.org/abs/1401.3299}{{\ttfamily 1401.3299}}].

\bibitem{Laiho:2016nlp}
J.~Laiho, S.~Bassler, D.~Coumbe, D.~Du and J.T.~Neelakanta, \emph{{Lattice
  Quantum Gravity and Asymptotic Safety}},
  \href{https://doi.org/10.1103/PhysRevD.96.064015}{\emph{Phys. Rev. D}
  {\bfseries 96} (2017) 064015}
  [\href{https://arxiv.org/abs/1604.02745}{{\ttfamily 1604.02745}}].

\bibitem{Ambjorn:1998xu}
J.~Ambjorn and R.~Loll, \emph{{Nonperturbative Lorentzian quantum gravity,
  causality and topology change}},
  \href{https://doi.org/10.1016/S0550-3213(98)00692-0}{\emph{Nucl. Phys. B}
  {\bfseries 536} (1998) 407}
  [\href{https://arxiv.org/abs/hep-th/9805108}{{\ttfamily hep-th/9805108}}].

\bibitem{Ambjorn:2001cv}
J.~Ambjorn, J.~Jurkiewicz and R.~Loll, \emph{{Dynamically triangulating
  Lorentzian quantum gravity}},
  \href{https://doi.org/10.1016/S0550-3213(01)00297-8}{\emph{Nucl. Phys. B}
  {\bfseries 610} (2001) 347}
  [\href{https://arxiv.org/abs/hep-th/0105267}{{\ttfamily hep-th/0105267}}].

\bibitem{Ambjorn:2005qt}
J.~Ambjorn, J.~Jurkiewicz and R.~Loll, \emph{{Reconstructing the universe}},
  \href{https://doi.org/10.1103/PhysRevD.72.064014}{\emph{Phys. Rev. D}
  {\bfseries 72} (2005) 064014}
  [\href{https://arxiv.org/abs/hep-th/0505154}{{\ttfamily hep-th/0505154}}].

\bibitem{Visser:2017atf}
M.~Visser, \emph{{How to Wick rotate generic curved spacetime}},
  \href{https://arxiv.org/abs/1702.05572}{{\ttfamily 1702.05572}}.

\bibitem{Baldazzi:2018mtl}
A.~Baldazzi, R.~Percacci and V.~Skrinjar, \emph{{Wicked metrics}},
  \href{https://doi.org/10.1088/1361-6382/ab187d}{\emph{Class. Quant. Grav.}
  {\bfseries 36} (2019) 105008}
  [\href{https://arxiv.org/abs/1811.03369}{{\ttfamily 1811.03369}}].

\bibitem{Ambjorn:2012ij}
J.~Ambjorn, S.~Jordan, J.~Jurkiewicz and R.~Loll, \emph{{Second- and
  First-Order Phase Transitions in CDT}},
  \href{https://doi.org/10.1103/PhysRevD.85.124044}{\emph{Phys. Rev. D}
  {\bfseries 85} (2012) 124044}
  [\href{https://arxiv.org/abs/1205.1229}{{\ttfamily 1205.1229}}].

\bibitem{Ambjorn:2012jv}
J.~Ambjorn, A.~Goerlich, J.~Jurkiewicz and R.~Loll, \emph{{Nonperturbative
  Quantum Gravity}},
  \href{https://doi.org/10.1016/j.physrep.2012.03.007}{\emph{Phys. Rept.}
  {\bfseries 519} (2012) 127}
  [\href{https://arxiv.org/abs/1203.3591}{{\ttfamily 1203.3591}}].

\bibitem{Loll:2019rdj}
R.~Loll, \emph{{Quantum Gravity from Causal Dynamical Triangulations: A
  Review}}, \href{https://doi.org/10.1088/1361-6382/ab57c7}{\emph{Class. Quant.
  Grav.} {\bfseries 37} (2020) 013002}
  [\href{https://arxiv.org/abs/1905.08669}{{\ttfamily 1905.08669}}].

\bibitem{Rovelli:1990ph}
C.~Rovelli, \emph{{What Is Observable in Classical and Quantum Gravity?}},
  \href{https://doi.org/10.1088/0264-9381/8/2/011}{\emph{Class. Quant. Grav.}
  {\bfseries 8} (1991) 297}.

\bibitem{Giddings:2005id}
S.B.~Giddings, D.~Marolf and J.B.~Hartle, \emph{{Observables in effective
  gravity}}, \href{https://doi.org/10.1103/PhysRevD.74.064018}{\emph{Phys. Rev.
  D} {\bfseries 74} (2006) 064018}
  [\href{https://arxiv.org/abs/hep-th/0512200}{{\ttfamily hep-th/0512200}}].

\bibitem{Ambjorn:2004pw}
J.~Ambjorn, J.~Jurkiewicz and R.~Loll, \emph{{Semiclassical universe from first
  principles}},
  \href{https://doi.org/10.1016/j.physletb.2004.12.067}{\emph{Phys. Lett. B}
  {\bfseries 607} (2005) 205}
  [\href{https://arxiv.org/abs/hep-th/0411152}{{\ttfamily hep-th/0411152}}].

\bibitem{Dasgupta:2001ue}
A.~Dasgupta and R.~Loll, \emph{{A Proper time cure for the conformal sickness
  in quantum gravity}},
  \href{https://doi.org/10.1016/S0550-3213(01)00227-9}{\emph{Nucl. Phys. B}
  {\bfseries 606} (2001) 357}
  [\href{https://arxiv.org/abs/hep-th/0103186}{{\ttfamily hep-th/0103186}}].

\bibitem{Horava:2008ih}
P.~Horava, \emph{{Membranes at Quantum Criticality}},
  \href{https://doi.org/10.1088/1126-6708/2009/03/020}{\emph{JHEP} {\bfseries
  03} (2009) 020} [\href{https://arxiv.org/abs/0812.4287}{{\ttfamily
  0812.4287}}].

\bibitem{Horava:2009uw}
P.~Horava, \emph{{Quantum Gravity at a Lifshitz Point}},
  \href{https://doi.org/10.1103/PhysRevD.79.084008}{\emph{Phys. Rev. D}
  {\bfseries 79} (2009) 084008}
  [\href{https://arxiv.org/abs/0901.3775}{{\ttfamily 0901.3775}}].

\bibitem{Wang:2017brl}
A.~Wang, \emph{{Ho\v{r}ava gravity at a Lifshitz point: A progress report}},
  \href{https://doi.org/10.1142/S0218271817300142}{\emph{Int. J. Mod. Phys. D}
  {\bfseries 26} (2017) 1730014}
  [\href{https://arxiv.org/abs/1701.06087}{{\ttfamily 1701.06087}}].

\bibitem{Steinwachs:2020jkj}
C.F.~Steinwachs, \emph{{Towards a unitary, renormalizable and
  ultraviolet-complete quantum theory of gravity}},
  \href{https://doi.org/10.3389/fphy.2020.00185}{\emph{Frontiers in Physics}
  {\bfseries 8} (2020) } [\href{https://arxiv.org/abs/2004.07842}{{\ttfamily
  2004.07842}}].

\bibitem{Horava:2009if}
P.~Horava, \emph{{Spectral Dimension of the Universe in Quantum Gravity at a
  Lifshitz Point}},
  \href{https://doi.org/10.1103/PhysRevLett.102.161301}{\emph{Phys. Rev. Lett.}
  {\bfseries 102} (2009) 161301}
  [\href{https://arxiv.org/abs/0902.3657}{{\ttfamily 0902.3657}}].

\bibitem{Benedetti:2009ge}
D.~Benedetti and J.~Henson, \emph{{Spectral geometry as a probe of quantum
  spacetime}}, \href{https://doi.org/10.1103/PhysRevD.80.124036}{\emph{Phys.
  Rev. D} {\bfseries 80} (2009) 124036}
  [\href{https://arxiv.org/abs/0911.0401}{{\ttfamily 0911.0401}}].

\bibitem{Ambjorn:2010hu}
J.~Ambjorn, A.~Gorlich, S.~Jordan, J.~Jurkiewicz and R.~Loll, \emph{{CDT meets
  Horava-Lifshitz gravity}},
  \href{https://doi.org/10.1016/j.physletb.2010.05.054}{\emph{Phys. Lett. B}
  {\bfseries 690} (2010) 413}
  [\href{https://arxiv.org/abs/1002.3298}{{\ttfamily 1002.3298}}].

\bibitem{Budd:2011zm}
T.G.~Budd, \emph{{The effective kinetic term in CDT}},
  \href{https://doi.org/10.1088/1742-6596/360/1/012038}{\emph{J. Phys. Conf.
  Ser.} {\bfseries 36} (2012) 012038}
  [\href{https://arxiv.org/abs/1110.5158}{{\ttfamily 1110.5158}}].

\bibitem{Ambjorn:2013joa}
J.~Ambj\o{}rn, L.~Glaser, Y.~Sato and Y.~Watabiki, \emph{{2d CDT is 2d
  Ho\v{r}ava\textendash{}Lifshitz quantum gravity}},
  \href{https://doi.org/10.1016/j.physletb.2013.04.006}{\emph{Phys. Lett. B}
  {\bfseries 722} (2013) 172}
  [\href{https://arxiv.org/abs/1302.6359}{{\ttfamily 1302.6359}}].

\bibitem{Benedetti:2014dra}
D.~Benedetti and J.~Henson, \emph{{Spacetime condensation in (2+1)-dimensional
  CDT from a Ho\v{r}ava\textendash{}Lifshitz minisuperspace model}},
  \href{https://doi.org/10.1088/0264-9381/32/21/215007}{\emph{Class. Quant.
  Grav.} {\bfseries 32} (2015) 215007}
  [\href{https://arxiv.org/abs/1410.0845}{{\ttfamily 1410.0845}}].

\bibitem{Benedetti:2016rwo}
D.~Benedetti and J.P.~Ryan, \emph{{Capturing the phase diagram of (2 +
  1)-dimensional CDT using a balls-in-boxes model}},
  \href{https://doi.org/10.1088/1361-6382/aa6b5d}{\emph{Class. Quant. Grav.}
  {\bfseries 34} (2017) 105012}
  [\href{https://arxiv.org/abs/1612.09533}{{\ttfamily 1612.09533}}].

\bibitem{Anderson:2011bj}
C.~Anderson, S.J.~Carlip, J.H.~Cooperman, P.~Horava, R.K.~Kommu and
  P.R.~Zulkowski, \emph{{Quantizing Horava-Lifshitz Gravity via Causal
  Dynamical Triangulations}},
  \href{https://doi.org/10.1103/PhysRevD.85.044027}{\emph{Phys. Rev. D}
  {\bfseries 85} (2012) 044027}
  [\href{https://arxiv.org/abs/1111.6634}{{\ttfamily 1111.6634}}].

\bibitem{Borji:2020snd}
M.~Borji and C.~Kopper, \emph{{Perturbative renormalization of the lattice
  regularized $\phi_4^4$ with flow equations}},
  \href{https://doi.org/10.1063/5.0024211}{\emph{J. Math. Phys.} {\bfseries 61}
  (2020) 112304} [\href{https://arxiv.org/abs/2006.15943}{{\ttfamily
  2006.15943}}].

\bibitem{Testa:1997ia}
M.~Testa, \emph{{The Rome approach to chirality}},  in \emph{{APCTP - ICTP
  Joint International Conference (AIJIC 97) on Recent Developments in
  Nonperturbative Quantum Field Theory}}, pp.~114--127, 7, 1997
  [\href{https://arxiv.org/abs/hep-lat/9707007}{{\ttfamily hep-lat/9707007}}].

\bibitem{Magnen:1992wv}
J.~Magnen, V.~Rivasseau and R.~Seneor, \emph{{Construction of Y M(4) with an
  infrared cutoff}}, \href{https://doi.org/10.1007/BF02097397}{\emph{Commun.
  Math. Phys.} {\bfseries 155} (1993) 325}.

\bibitem{Ellwanger:1994iz}
U.~Ellwanger, \emph{{Flow equations and BRS invariance for Yang-Mills
  theories}}, \href{https://doi.org/10.1016/0370-2693(94)90365-4}{\emph{Phys.
  Lett. B} {\bfseries 335} (1994) 364}
  [\href{https://arxiv.org/abs/hep-th/9402077}{{\ttfamily hep-th/9402077}}].

\bibitem{Gies:2006wv}
H.~Gies, \emph{{Introduction to the functional RG and applications to gauge
  theories}}, \href{https://doi.org/10.1007/978-3-642-27320-9_6}{\emph{Lect.
  Notes Phys.} {\bfseries 852} (2012) 287}
  [\href{https://arxiv.org/abs/hep-ph/0611146}{{\ttfamily hep-ph/0611146}}].

\bibitem{Goldenfeld:1992qy}
N.~Goldenfeld, \emph{{Lectures on phase transitions and the renormalization
  group}}, CRC Press (1992).

\bibitem{Johnston:1994qi}
D.A.~Johnston, J.P.~Kownacki and A.~Krzywicki, \emph{{Random geometries and
  real space renormalization group}},
  \href{https://doi.org/10.1016/0920-5632(95)00364-F}{\emph{Nucl. Phys. B Proc.
  Suppl.} {\bfseries 42} (1995) 728}
  [\href{https://arxiv.org/abs/hep-lat/9407018}{{\ttfamily hep-lat/9407018}}].

\bibitem{Thorleifsson:1995ki}
G.~Thorleifsson and S.~Catterall, \emph{{A real space renormalization group for
  random surfaces}},
  \href{https://doi.org/10.1016/0550-3213(95)00664-8}{\emph{Nucl. Phys. B}
  {\bfseries 461} (1996) 350}
  [\href{https://arxiv.org/abs/hep-lat/9510003}{{\ttfamily hep-lat/9510003}}].

\bibitem{Ambjorn:1996hu}
J.~Ambjorn, P.~Bialas and J.~Jurkiewicz, \emph{{RG flow in an exactly solvable
  model with fluctuating geometry}},
  \href{https://doi.org/10.1016/0370-2693(96)00457-1}{\emph{Phys. Lett. B}
  {\bfseries 379} (1996) 93}
  [\href{https://arxiv.org/abs/hep-lat/9602021}{{\ttfamily hep-lat/9602021}}].

\bibitem{Renken:1996kf}
R.L.~Renken, \emph{{A Renormalization group for dynamical triangulations in
  arbitrary dimensions}},
  \href{https://doi.org/10.1016/S0550-3213(96)00611-6}{\emph{Nucl. Phys. B}
  {\bfseries 485} (1997) 503}
  [\href{https://arxiv.org/abs/hep-lat/9607074}{{\ttfamily hep-lat/9607074}}].

\bibitem{Henson:2009fy}
J.~Henson, \emph{{Coarse graining dynamical triangulations: A New scheme}},
  \href{https://doi.org/10.1088/0264-9381/26/17/175019}{\emph{Class. Quant.
  Grav.} {\bfseries 26} (2009) 175019}
  [\href{https://arxiv.org/abs/0907.5602}{{\ttfamily 0907.5602}}].

\bibitem{Markopoulou:2002ja}
F.~Markopoulou, \emph{{Coarse graining in spin foam models}},
  \href{https://doi.org/10.1088/0264-9381/20/5/301}{\emph{Class. Quant. Grav.}
  {\bfseries 20} (2003) 777}
  [\href{https://arxiv.org/abs/gr-qc/0203036}{{\ttfamily gr-qc/0203036}}].

\bibitem{Oeckl:2002ia}
R.~Oeckl, \emph{{Renormalization of discrete models without background}},
  \href{https://doi.org/10.1016/S0550-3213(03)00145-7}{\emph{Nucl. Phys. B}
  {\bfseries 657} (2003) 107}
  [\href{https://arxiv.org/abs/gr-qc/0212047}{{\ttfamily gr-qc/0212047}}].

\bibitem{Bahr:2012qj}
B.~Bahr, B.~Dittrich, F.~Hellmann and W.~Kaminski, \emph{{Holonomy Spin Foam
  Models: Definition and Coarse Graining}},
  \href{https://doi.org/10.1103/PhysRevD.87.044048}{\emph{Phys. Rev. D}
  {\bfseries 87} (2013) 044048}
  [\href{https://arxiv.org/abs/1208.3388}{{\ttfamily 1208.3388}}].

\bibitem{Dittrich:2013uqe}
B.~Dittrich, M.~Mart\'\i{}n-Benito and E.~Schnetter, \emph{{Coarse graining of
  spin net models: dynamics of intertwiners}},
  \href{https://doi.org/10.1088/1367-2630/15/10/103004}{\emph{New J. Phys.}
  {\bfseries 15} (2013) 103004}
  [\href{https://arxiv.org/abs/1306.2987}{{\ttfamily 1306.2987}}].

\bibitem{Steinhaus:2020lgb}
S.~Steinhaus, \emph{{Coarse Graining Spin Foam Quantum Gravity\textemdash{}A
  Review}}, \href{https://doi.org/10.3389/fphy.2020.00295}{\emph{Front. in
  Phys.} {\bfseries 8} (2020) 295}
  [\href{https://arxiv.org/abs/2007.01315}{{\ttfamily 2007.01315}}].

\bibitem{Renken:1997na}
R.L.~Renken, S.M.~Catterall and J.B.~Kogut, \emph{{Phase structure of dynamical
  triangulation models in three-dimensions}},
  \href{https://doi.org/10.1016/S0550-3213(98)00142-4}{\emph{Nucl. Phys. B}
  {\bfseries 523} (1998) 553}
  [\href{https://arxiv.org/abs/hep-lat/9712011}{{\ttfamily hep-lat/9712011}}].

\bibitem{Bialas:1998ci}
P.~Bialas, Z.~Burda and D.~Johnston, \emph{{Phase diagram of the mean field
  model of simplicial gravity}},
  \href{https://doi.org/10.1016/S0550-3213(98)00842-6}{\emph{Nucl. Phys. B}
  {\bfseries 542} (1999) 413}
  [\href{https://arxiv.org/abs/gr-qc/9808011}{{\ttfamily gr-qc/9808011}}].

\bibitem{Ambjorn:2001br}
J.~Ambjorn, J.~Jurkiewicz, R.~Loll and G.~Vernizzi, \emph{{Lorentzian 3-D
  gravity with wormholes via matrix models}},
  \href{https://doi.org/10.1088/1126-6708/2001/09/022}{\emph{JHEP} {\bfseries
  09} (2001) 022} [\href{https://arxiv.org/abs/hep-th/0106082}{{\ttfamily
  hep-th/0106082}}].

\bibitem{Evans-review}
M.R.~Evans and T.~Hanney, \emph{Nonequilibrium statistical mechanics of the
  zero-range process and related models}, {\emph{J. Phys. A: Math. and Gen.}
  {\bfseries 38} (2005) R195}.

\bibitem{Bialas:1996eh}
P.~Bialas, Z.~Burda, B.~Petersson and J.~Tabaczek, \emph{{Appearance of mother
  universe and singular vertices in random geometries}},
  \href{https://doi.org/10.1016/S0550-3213(97)00226-5}{\emph{Nucl. Phys. B}
  {\bfseries 495} (1997) 463}
  [\href{https://arxiv.org/abs/hep-lat/9608030}{{\ttfamily hep-lat/9608030}}].

\bibitem{Bialas:1997qs}
P.~Bialas, Z.~Burda and D.~Johnston, \emph{{Condensation in the backgammon
  model}}, \href{https://doi.org/10.1016/S0550-3213(97)00192-2}{\emph{Nucl.
  Phys. B} {\bfseries 493} (1997) 505}
  [\href{https://arxiv.org/abs/cond-mat/9609264}{{\ttfamily
  cond-mat/9609264}}].

\bibitem{Bogacz:2012sa}
L.~Bogacz, Z.~Burda and B.~Waclaw, \emph{{Quantum widening of CDT universe}},
  \href{https://doi.org/10.1103/PhysRevD.86.104015}{\emph{Phys. Rev. D}
  {\bfseries 86} (2012) 104015}
  [\href{https://arxiv.org/abs/1204.1356}{{\ttfamily 1204.1356}}].

\bibitem{Evans-prl}
M.R.~Evans, T.~Hanney and S.N.~Majumdar, \emph{Interaction-driven real-space
  condensation},
  \href{https://doi.org/10.1103/PhysRevLett.97.010602}{\emph{Phys. Rev. Lett.}
  {\bfseries 97} (2006) 010602}.

\bibitem{Waclaw:2009zz}
B.~Waclaw, J.~Sopik, W.~Janke and H.~Meyer-Ortmanns, \emph{{Tuning the shape of
  the condensate in spontaneous symmetry breaking}},
  \href{https://doi.org/10.1103/PhysRevLett.103.080602}{\emph{Phys. Rev. Lett.}
  {\bfseries 103} (2009) 080602}
  [\href{https://arxiv.org/abs/0901.3664}{{\ttfamily 0901.3664}}].

\bibitem{DiFrancesco:1999em}
P.~Di~Francesco, E.~Guitter and C.~Kristjansen, \emph{{Integrable 2-D
  Lorentzian gravity and random walks}},
  \href{https://doi.org/10.1016/S0550-3213(99)00661-6}{\emph{Nucl. Phys. B}
  {\bfseries 567} (2000) 515}
  [\href{https://arxiv.org/abs/hep-th/9907084}{{\ttfamily hep-th/9907084}}].

\bibitem{DiFrancesco:2000nn}
P.~Di~Francesco, E.~Guitter and C.~Kristjansen, \emph{{Generalized Lorentzian
  triangulations and the Calogero Hamiltonian}},
  \href{https://doi.org/10.1016/S0550-3213(01)00239-5}{\emph{Nucl. Phys. B}
  {\bfseries 608} (2001) 485}
  [\href{https://arxiv.org/abs/hep-th/0010259}{{\ttfamily hep-th/0010259}}].

\bibitem{DiFrancesco:2001xur}
P.~Di~Francesco and E.~Guitter, \emph{{Critical and multicritical semirandom
  (1+d)-dimensional lattices and hard objects in d-dimensions}},
  \href{https://doi.org/10.1088/0305-4470/35/4/304}{\emph{J. Phys. A}
  {\bfseries 35} (2002) 897}
  [\href{https://arxiv.org/abs/cond-mat/0104383}{{\ttfamily
  cond-mat/0104383}}].

\bibitem{Murtazaev:2013}
A.K.~Murtazaev and Z.G.~Ibaev, \emph{On choosing the order parameter of
  modulated magnetic structures}, {\emph{Journal of Experimental and
  Theoretical Physics} {\bfseries 116} (2013) 266}.

\bibitem{Sato:2022ory}
Y.~Sato, \emph{{CDT and Horava-Lifshitz QG in Two Dimensions}},  in
  \emph{Handbook of Quantum Gravity}, C.~Bambi, L.~Modesto and I.~Shapiro,
  eds., Springer Singapore (2023)
  [\href{https://arxiv.org/abs/2212.03446}{{\ttfamily 2212.03446}}].

\bibitem{Mattei:2005cm}
F.~Mattei, C.~Rovelli, S.~Speziale and M.~Testa, \emph{{From 3-geometry
  transition amplitudes to graviton states}},
  \href{https://doi.org/10.1016/j.nuclphysb.2006.01.026}{\emph{Nucl. Phys. B}
  {\bfseries 739} (2006) 234}
  [\href{https://arxiv.org/abs/gr-qc/0508007}{{\ttfamily gr-qc/0508007}}].

\bibitem{Ambjorn:1999gi}
J.~Ambjorn, K.N.~Anagnostopoulos and R.~Loll, \emph{{A New perspective on
  matter coupling in 2-D quantum gravity}},
  \href{https://doi.org/10.1103/PhysRevD.60.104035}{\emph{Phys. Rev. D}
  {\bfseries 60} (1999) 104035}
  [\href{https://arxiv.org/abs/hep-th/9904012}{{\ttfamily hep-th/9904012}}].

\bibitem{Ambjorn:2015gea}
J.~Ambj\o{}rn, A.~G\"orlich, J.~Jurkiewicz and H.~Zhang, \emph{{The microscopic
  structure of 2D CDT coupled to matter}},
  \href{https://doi.org/10.1016/j.physletb.2015.05.026}{\emph{Phys. Lett. B}
  {\bfseries 746} (2015) 359}
  [\href{https://arxiv.org/abs/1503.01636}{{\ttfamily 1503.01636}}].

\bibitem{Benedetti:2007pp}
D.~Benedetti, R.~Loll and F.~Zamponi, \emph{{(2+1)-dimensional quantum gravity
  as the continuum limit of Causal Dynamical Triangulations}},
  \href{https://doi.org/10.1103/PhysRevD.76.104022}{\emph{Phys. Rev. D}
  {\bfseries 76} (2007) 104022}
  [\href{https://arxiv.org/abs/0704.3214}{{\ttfamily 0704.3214}}].

\bibitem{Durhuus:2014dbl}
B.~Durhuus and T.~Jonsson, \emph{{Exponential Bounds on the Number of Causal
  Triangulations}},
  \href{https://doi.org/10.1007/s00220-015-2453-2}{\emph{Commun. Math. Phys.}
  {\bfseries 340} (2015) 105}
  [\href{https://arxiv.org/abs/1408.2101}{{\ttfamily 1408.2101}}].

\bibitem{Ambjorn:2000dja}
J.~Ambjorn, J.~Jurkiewicz and R.~Loll, \emph{{Nonperturbative 3-D Lorentzian
  quantum gravity}},
  \href{https://doi.org/10.1103/PhysRevD.64.044011}{\emph{Phys. Rev. D}
  {\bfseries 64} (2001) 044011}
  [\href{https://arxiv.org/abs/hep-th/0011276}{{\ttfamily hep-th/0011276}}].

\bibitem{Cooperman:2013mma}
J.H.~Cooperman and J.~Miller, \emph{{A first look at transition amplitudes in
  (2 + 1)-dimensional causal dynamical triangulations}},
  \href{https://doi.org/10.1088/0264-9381/31/3/035012}{\emph{Class. Quant.
  Grav.} {\bfseries 31} (2014) 035012}
  [\href{https://arxiv.org/abs/1305.2932}{{\ttfamily 1305.2932}}].

\bibitem{Budd:2013waa}
T.G.~Budd and R.~Loll, \emph{{Exploring Torus Universes in Causal Dynamical
  Triangulations}},
  \href{https://doi.org/10.1103/PhysRevD.88.024015}{\emph{Phys. Rev. D}
  {\bfseries 88} (2013) 024015}
  [\href{https://arxiv.org/abs/1305.4702}{{\ttfamily 1305.4702}}].

\bibitem{Ambjorn:2008wc}
J.~Ambjorn, A.~Gorlich, J.~Jurkiewicz and R.~Loll, \emph{{The Nonperturbative
  Quantum de Sitter Universe}},
  \href{https://doi.org/10.1103/PhysRevD.78.063544}{\emph{Phys. Rev. D}
  {\bfseries 78} (2008) 063544}
  [\href{https://arxiv.org/abs/0807.4481}{{\ttfamily 0807.4481}}].

\bibitem{Gorlich:2011ga}
A.~Gorlich, \emph{{Causal Dynamical Triangulations in Four Dimensions}}, Ph.D.
  thesis, Jagiellonian U., Astron. Observ., 2010.
\newblock \href{https://arxiv.org/abs/1111.6938}{{\ttfamily 1111.6938}}.

\bibitem{Ambjorn:2007jv}
J.~Ambjorn, A.~Gorlich, J.~Jurkiewicz and R.~Loll, \emph{{Planckian Birth of
  the Quantum de Sitter Universe}},
  \href{https://doi.org/10.1103/PhysRevLett.100.091304}{\emph{Phys. Rev. Lett.}
  {\bfseries 100} (2008) 091304}
  [\href{https://arxiv.org/abs/0712.2485}{{\ttfamily 0712.2485}}].

\bibitem{Ambjorn:2011ph}
J.~Ambjorn, A.~Gorlich, J.~Jurkiewicz, R.~Loll, J.~Gizbert-Studnicki and
  T.~Trzesniewski, \emph{{The Semiclassical Limit of Causal Dynamical
  Triangulations}},
  \href{https://doi.org/10.1016/j.nuclphysb.2011.03.019}{\emph{Nucl. Phys. B}
  {\bfseries 849} (2011) 144}
  [\href{https://arxiv.org/abs/1102.3929}{{\ttfamily 1102.3929}}].

\bibitem{Ambjorn:2012pp}
J.~Ambjorn, J.~Gizbert-Studnicki, A.~Gorlich and J.~Jurkiewicz, \emph{{The
  Transfer matrix in four-dimensional CDT}},
  \href{https://doi.org/10.1007/JHEP09(2012)017}{\emph{JHEP} {\bfseries 09}
  (2012) 017} [\href{https://arxiv.org/abs/1205.3791}{{\ttfamily 1205.3791}}].

\bibitem{Gibbons:1978ac}
G.W.~Gibbons, S.W.~Hawking and M.J.~Perry, \emph{{Path Integrals and the
  Indefiniteness of the Gravitational Action}},
  \href{https://doi.org/10.1016/0550-3213(78)90161-X}{\emph{Nucl. Phys. B}
  {\bfseries 138} (1978) 141}.

\bibitem{Hartle:1983ai}
J.B.~Hartle and S.W.~Hawking, \emph{{Wave Function of the Universe}},
  \href{https://doi.org/10.1103/PhysRevD.28.2960}{\emph{Phys. Rev. D}
  {\bfseries 28} (1983) 2960}.

\bibitem{Ambjorn:2014mra}
J.~Ambj\o{}rn, J.~Gizbert-Studnicki, A.~G\"orlich and J.~Jurkiewicz, \emph{{The
  effective action in 4-dim CDT. The transfer matrix approach}},
  \href{https://doi.org/10.1007/JHEP06(2014)034}{\emph{JHEP} {\bfseries 06}
  (2014) 034} [\href{https://arxiv.org/abs/1403.5940}{{\ttfamily 1403.5940}}].

\bibitem{Ambjorn:2016mnn}
J.~Ambj\o{}rn, J.~Gizbert-Studnicki, A.~G\"orlich, J.~Jurkiewicz, N.~Klitgaard
  and R.~Loll, \emph{{Characteristics of the new phase in CDT}},
  \href{https://doi.org/10.1140/epjc/s10052-017-4710-3}{\emph{Eur. Phys. J. C}
  {\bfseries 77} (2017) 152}
  [\href{https://arxiv.org/abs/1610.05245}{{\ttfamily 1610.05245}}].

\bibitem{Ambjorn:2002nu}
J.~Ambjorn, J.~Jurkiewicz and R.~Loll, \emph{{3-d Lorentzian, dynamically
  triangulated quantum gravity}},
  \href{https://doi.org/10.1016/S0920-5632(01)01904-1}{\emph{Nucl. Phys. B
  Proc. Suppl.} {\bfseries 106} (2002) 980}
  [\href{https://arxiv.org/abs/hep-lat/0201013}{{\ttfamily hep-lat/0201013}}].

\bibitem{Ambjorn:2014gsa}
J.~Ambjorn, A.~G\"orlich, J.~Jurkiewicz, A.~Kreienbuehl and R.~Loll,
  \emph{{Renormalization Group Flow in CDT}},
  \href{https://doi.org/10.1088/0264-9381/31/16/165003}{\emph{Class. Quant.
  Grav.} {\bfseries 31} (2014) 165003}
  [\href{https://arxiv.org/abs/1405.4585}{{\ttfamily 1405.4585}}].

\bibitem{Ambjorn:2020rcn}
J.~Ambjorn, J.~Gizbert-Studnicki, A.~G\"orlich, J.~Jurkiewicz and R.~Loll,
  \emph{{Renormalization in quantum theories of geometry}},
  \href{https://doi.org/10.3389/fphy.2020.00247}{\emph{Front. in Phys.}
  {\bfseries 8} (2020) 247} [\href{https://arxiv.org/abs/2002.01693}{{\ttfamily
  2002.01693}}].

\bibitem{Jordan:2013awa}
S.~Jordan and R.~Loll, \emph{{Causal Dynamical Triangulations without Preferred
  Foliation}},
  \href{https://doi.org/10.1016/j.physletb.2013.06.007}{\emph{Phys. Lett. B}
  {\bfseries 724} (2013) 155}
  [\href{https://arxiv.org/abs/1305.4582}{{\ttfamily 1305.4582}}].

\bibitem{Jordan:2013iaa}
S.~Jordan and R.~Loll, \emph{{De Sitter Universe from Causal Dynamical
  Triangulations without Preferred Foliation}},
  \href{https://doi.org/10.1103/PhysRevD.88.044055}{\emph{Phys. Rev. D}
  {\bfseries 88} (2013) 044055}
  [\href{https://arxiv.org/abs/1307.5469}{{\ttfamily 1307.5469}}].

\bibitem{Ambjorn:2005db}
J.~Ambjorn, J.~Jurkiewicz and R.~Loll, \emph{{Spectral dimension of the
  universe}}, \href{https://doi.org/10.1103/PhysRevLett.95.171301}{\emph{Phys.
  Rev. Lett.} {\bfseries 95} (2005) 171301}
  [\href{https://arxiv.org/abs/hep-th/0505113}{{\ttfamily hep-th/0505113}}].

\bibitem{Lauscher:2005qz}
O.~Lauscher and M.~Reuter, \emph{{Fractal spacetime structure in asymptotically
  safe gravity}},
  \href{https://doi.org/10.1088/1126-6708/2005/10/050}{\emph{JHEP} {\bfseries
  10} (2005) 050} [\href{https://arxiv.org/abs/hep-th/0508202}{{\ttfamily
  hep-th/0508202}}].

\bibitem{Benedetti:2013pya}
D.~Benedetti and F.~Guarnieri, \emph{{One-loop renormalization in a toy model
  of Ho\v{r}ava-Lifshitz gravity}},
  \href{https://doi.org/10.1007/JHEP03(2014)078}{\emph{JHEP} {\bfseries 03}
  (2014) 078} [\href{https://arxiv.org/abs/1311.6253}{{\ttfamily 1311.6253}}].

\bibitem{Barvinsky:2017kob}
A.O.~Barvinsky, D.~Blas, M.~Herrero-Valea, S.M.~Sibiryakov and C.F.~Steinwachs,
  \emph{{Ho\v{r}ava Gravity is Asymptotically Free in 2 + 1 Dimensions}},
  \href{https://doi.org/10.1103/PhysRevLett.119.211301}{\emph{Phys. Rev. Lett.}
  {\bfseries 119} (2017) 211301}
  [\href{https://arxiv.org/abs/1706.06809}{{\ttfamily 1706.06809}}].

\bibitem{Barvinsky:2019rwn}
A.O.~Barvinsky, M.~Herrero-Valea and S.M.~Sibiryakov, \emph{{Towards the
  renormalization group flow of Horava gravity in $(3+1)$ dimensions}},
  \href{https://doi.org/10.1103/PhysRevD.100.026012}{\emph{Phys. Rev. D}
  {\bfseries 100} (2019) 026012}
  [\href{https://arxiv.org/abs/1905.03798}{{\ttfamily 1905.03798}}].

\bibitem{Barvinsky:2021ubv}
A.O.~Barvinsky, A.V.~Kurov and S.M.~Sibiryakov, \emph{{Beta functions of
  (3+1)-dimensional projectable Ho\v{r}ava gravity}},
  \href{https://doi.org/10.1103/PhysRevD.105.044009}{\emph{Phys. Rev. D}
  {\bfseries 105} (2022) 044009}
  [\href{https://arxiv.org/abs/2110.14688}{{\ttfamily 2110.14688}}].

\bibitem{Benedetti:2013jk}
D.~Benedetti, \emph{{On the number of relevant operators in asymptotically safe
  gravity}}, \href{https://doi.org/10.1209/0295-5075/102/20007}{\emph{EPL}
  {\bfseries 102} (2013) 20007}
  [\href{https://arxiv.org/abs/1301.4422}{{\ttfamily 1301.4422}}].

\bibitem{Mitchell:2021qjr}
A.~Mitchell, T.R.~Morris and D.~Stulga, \emph{{Provable properties of
  asymptotic safety in f(R) approximation}},
  \href{https://doi.org/10.1007/JHEP01(2022)041}{\emph{JHEP} {\bfseries 01}
  (2022) 041} [\href{https://arxiv.org/abs/2111.05067}{{\ttfamily
  2111.05067}}].

\bibitem{Codello:2008vh}
A.~Codello, R.~Percacci and C.~Rahmede, \emph{{Investigating the Ultraviolet
  Properties of Gravity with a Wilsonian Renormalization Group Equation}},
  \href{https://doi.org/10.1016/j.aop.2008.08.008}{\emph{Annals Phys.}
  {\bfseries 324} (2009) 414}
  [\href{https://arxiv.org/abs/0805.2909}{{\ttfamily 0805.2909}}].

\bibitem{Benedetti:2009rx}
D.~Benedetti, P.F.~Machado and F.~Saueressig, \emph{{Asymptotic safety in
  higher-derivative gravity}},
  \href{https://doi.org/10.1142/S0217732309031521}{\emph{Mod. Phys. Lett. A}
  {\bfseries 24} (2009) 2233}
  [\href{https://arxiv.org/abs/0901.2984}{{\ttfamily 0901.2984}}].

\bibitem{Falls:2014tra}
K.~Falls, D.F.~Litim, K.~Nikolakopoulos and C.~Rahmede, \emph{{Further evidence
  for asymptotic safety of quantum gravity}},
  \href{https://doi.org/10.1103/PhysRevD.93.104022}{\emph{Phys. Rev. D}
  {\bfseries 93} (2016) 104022}
  [\href{https://arxiv.org/abs/1410.4815}{{\ttfamily 1410.4815}}].

\end{thebibliography}
\end{document}